# Determination of some solubilization parameters with surfactants of egg-yolk lecithin multilamellar vesicles by static light–scattering measurements.

ŞTEFAN HOBAI, ZITA FAZAKAS
*Department of Biochemistry, University of Medicine and Pharmacy, Tg. Mureş, Romania*

Abstract

Effective surfactant:phospholipid ratios (i.e. molar ratios in the mixed aggregates, vesicles or micelles) have been determined by static light-scattering for the interaction of egg-yolk lecithin (EYL) multilamellar vesicles (MLV) with Triton X-100 (TX-100), sodium deoxycholate (DOCNa) and cetyltrimethylammonium bromide (CTMB). The suspension of MLV-EYL was mixed with appropriate volumes of surfactant solution and was left overnight to reaches thermodynamic equilibrium. Rectangular optic diffusion data were used to compute the solubilization parameters: total surfactant concentrations, at saturation and solubilization $D_t^{sat}$ and $D_t^{sol}$ respectively, and effective molar ratios, $R_e^{sat}$ and $R_e^{sol}$ respectively. From the $R_e^{sat}$ value obtained graphically for interaction of vesicles with TX-100 resulted that in vesicle bilayers a surfactant molecule is surrounded with seven phospholipid molecules and the $R_e^{sol}$ value suggests that in mixed micelles ten lipid molecules with about fifteen surfactant molecules coexist. The values of $R_e^{sat}$ and $R_e^{sol}$ in case of MLV solubilization with DOCNa are 0.07 and 1.1 respectively. In case of vesicles solubilization with CTMB, $R_e^{sat}$ =0.25 and $R_e^{sol}$ =3.6 These data show that the solubilization power at thermodynamic equilibrium for the three surfactants is in the sequence: CTMB<TX-100<DOCNa.

## Introduction

Surfactants are widely used as molecular tools in membrane biochemistry and liposome-mediated drug delivery. Consequently, the interaction between lipid bilayers and surfactants has been the object of considerable attention [ Helenius, 1975; Lichtenberg , 1983, 1993, 1996; Cabantchik, 1985; Lasch, 1993, 1995; Fazakas, 1999]. Lichtenberg and co-workers [Schurtenberg, 1985;Almog, 1990; Lichtenberg, 1995] have established the guidelines for improved quantitative studies on membrane-surfactant interaction by defining the so-called "effective surfactant / lipid ratio", $R_e$. $R_e$ is defined as the surfactant / lipid molar ratio in the mixed aggregates, vesicles or micelles.

*Abbreviations:*
CMC, critical micellar concetration; CTMB, cetyltrimethylammonium bromide; Dif%, rectangular optical diffusion (percentage); $D_t^{sat}$, total surfactant concentration at saturation of vesicles; $D_t^{sol}$, total surfactant concentration necessary of vesicle solubilization; $D_w^{sat}$, concentration of surfactant in aqueous environment at bilayers saturation; $D_w^{sol}$, concentration of surfactant in aqueous environment at end of vesicle solubilization; DOCNa, sodium deoxycholate; EYL, egg-yolk lecithin; LUV, large unilamellar vesicles; MLV, multilamellar vesicles; $R_e$, effective molar ratio surfactant/phospholipid; $R_e^{sat}$, $R_e$ at the onset of solubilization; $R_e^{sol}$, $R_e$ at complete of solubilization; SUV, small unilamellar vesicles; TX-100, Triton X-100.

*Correspondence address:*
Ştefan Hobai, Department of Biochemistry, University of Medicine and Pharmacy, Târgu Mureş, Romania.
Fax:040-065-210407, e-mail: fazy@netsoft.ro.

The surfactant-induced transition from vesicles to mixed micelles, originally described by a three-staged model [Helenius, 1975; Lichtenberg, 1983; Schubert, 1986] has to be extended into a four-stage model. The sequence of events in the latter are as follows [Lasch,1995]. In stage I the surfactant distributes between water and the lipid phase resulting in mixed bilayer vesicles. In stage II mixed bilayer vesicles coexist with mixed bilayer sheets [Ollivon, 1988; Walter, 1991]. When a critical value of $R_e$ is reached ("saturation" of the bilayer), mixed bilayer transform progressively into lipid-rich mixed micelles (stage III) until all bilayers have disappeared in stage IV. In stage IV micelles grow gradually richer in surfactant with a concomitant decrease in particle size.

The initial distribution of surfactant may involve only the outer monolayer of the bilayer [Ueno, 1984]. In several cases the membrane / water partition coefficient describing the partitioning of small amounts of a surfactant appears to be a decreasing function of surfactant concentration [Schubert, 1986; Hobai, 1999]. Over a large range of sub-lytic surfactant concentrations, the partition of surfactant into membranes obeys partition coefficient:

$$K = D_b / D_w (L + D_b) \qquad (1)$$

where $D_b$ and $D_w$ are concentrations of surfactant contained in mixed vesicles and aqueous media, respectively, and L is lipid concentration. Knowing the value of K, the total surfactant concentra-tion, $D_t$, needed for a system with phospholipid concentration L to have an effective ratio $R_e$ [Almog, 1990] is given by expression:

$$D_t = R_e \{L + 1 / [ K ( R_e + 1 ) ] \} \qquad (2)$$

In stage I, $R_e \ll 1$ and equation (2) becomes:

$$D_t = R_e [L + (1 / K) ] \qquad (3)$$

The equations (2) and (3) mean that $D_t$, required for obtaining any effective ratio $R_e$, has a linear dependence on L. $R_e$ is given by the slope of the straight line defined by equation (2), which should intercept with the x-axis at $-1 / [ K ( R_e + 1 ) ]$.

Solubilization can also be simply followed by static light-scattering at 90° (rectangular optic diffusion). Light-scattering evolution during the solubilization depends on vesicle type [Partearroyo, 1992]. For small or large unilamellar vesicles (SUV or LUV) within the vesicular range surfactant-induced growth result in increased light-scattering. This increase of light scattering not appear in case of multilamellar vesicles (MLV). For SUV, LUV and MLV within the micellar range, increasing $D_t$ (thus $R_e$) results in a decrease in micellar size and a consequent decrease of light-scattering. Within the coexistence range increasing $R_e$ results in solubilization of a increasingly larger fraction of vesicles, thus in a consequent decrease of light-scattering. The surfactant concentration required for the onset of solubilization ($D_t^{sat}$) and the surfactant concentration required for complete solubilization ($D_t^{sol}$) as functions of L graphically are straight lines (eq. 2) of which slopes are $R_e^{sat}$ and $R_e^{sol}$, respectively. The intercepts $D_w^{sat}$ and $D_w^{sol}$ are critical micellar concentrations (CMC).

**Apparatus, materials and method**

Egg -yolk lecithine was purchased from Sigma Chemical Co. type X-E (cat.no. P5394/ 1996) purified by us by neutral alumina column chromatography, verified by thin layer chromatography (silicagel). The mobile phase was a mixture $CHCl_3:CH_3OH:H_2O$ (65:30:4,vol). The identification of the phospholipid was made with iodine and shows a single spot. Sodium deoxycholate (DOCNa) was purchased from Merck, the Triton X-100 (TX-100) from Sigma Chemical Co. and cetyltrimethylammonium bromide (CTMB) from Fluka AG. Tris(hydroxymethyl)aminomethane-HCl (Tris-Cl) was purchased from Austranal. All chemicals and organic solvents were of analytical and of spectroscopic grades, respectively.

By using a spectrophotometer Spekol with FR optical diffusion system and photocells and voltage source type HQE 40 and UV lamp type OSRAM, was determined the rectangular optic diffusion at λ=437 nm of MLV-egg-yolk lecithin suspension in the course of titration at 22.5°C with TX-100 20mM (Table 1), with DOCNa (Table 2) and with CTMB 20mM (Table 3). The sample was 1.4ml containing 0.5milimols lecithin and the titration rate was 0.08micromols surfactant/sec. At equilibrium condition the vesicle-surfactant mixtures were left overnight at room temperature.

**Liposome dispersion preparation**

Appropriate amount of EYL in chloroform:methanol (9:1,vol) solution was evaporated, in a glass test tube, under a nitrogen stream until it reached the consistence of a lipid film on the wall tube settled down. The film was purged with the gas for 30 minutes in order to eliminate the alcohol traces. The film hydration and bilayer suspension were made with buffer Tris-Cl 0.05M, pH=7.2 at 0.6mg/ml (lipid/solution) by vortexing 10 s.

**Results and Discussion**

The titration of MLV-EYL suspension with surfactants solutions was made in the rectangular optic diffusion device described above.

Table 1. *Rectangular optical diffusion (percentage), Dif%, of MLV-EYL in the course of titration with TX-100 20mM.*

| TX-100 volume, μl | log[TX-100] | Dif% |
|---|---|---|
| 10 | -3.85 | 97.53 |
| 20 | -3.55 | 95.06 |
| 30 | -3.38 | 86.42 |
| 40 | -3.25 | 80.25 |
| 60 | -3.09 | 66.67 |
| 80 | -2.97 | 49.38 |
| 100 | -2.88 | 43.21 |
| 120 | -2.80 | 37.04 |
| 150 | -2.71 | 34.57 |
| 200 | -2.6 | 32.72 |
| 300 | -2.45 | 31.48 |

Figure 1 shows the exploring data (solid rhombus) of total concentration of TX-100 (Table 1) used for the determination of parameters $R_e^{sat}$ and $R_e^{sol}$ at thermodynamic equilibrium.

**Figure 1.** *Titration curve of MLV-EYL suspension with TX-100. The solid rhombs correspond to the surfactant concentrations selected to be used to diffusion measurements at thermodynamic equilibrium.*

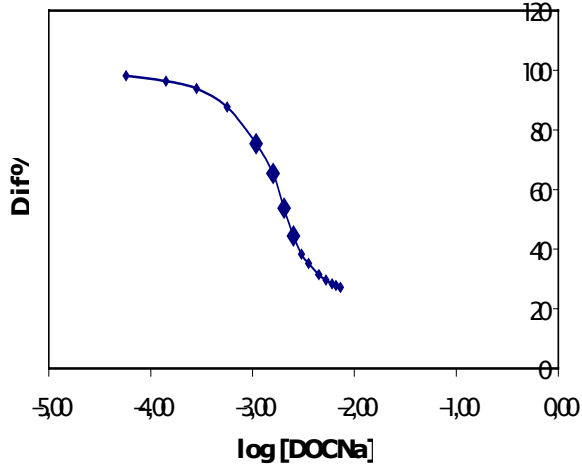

**Table 2.** *Rectangular optical diffusion (percentage), Dif%, of MLV-EYL in the course of titration with DOCNa 20mM.*

| DOCNa volume, µl | log[DOCNa] | Dif% |
|---|---|---|
| 4 | -4.24 | 98.15 |
| 10 | -3.85 | 96.3 |
| 20 | -3.55 | 93.83 |
| 40 | -3.25 | 87.66 |
| 80 | -2.97 | 75.31 |
| 120 | -2.8 | 65.44 |
| 160 | -2.69 | 53.09 |
| 200 | -2.60 | 44.45 |
| 250 | -2.52 | 38.27 |
| 300 | -2.45 | 35.19 |
| 400 | -2.35 | 31.48 |
| 500 | -2.28 | 29.63 |
| 600 | -2.22 | 28.4 |
| 700 | -2.18 | 27.78 |
| 800 | -2.14 | 27.16 |

Table 2 presents the experimental data of titration with DOCNa 20mM.

**Figure 2.** *Titration curve of MLV-EYL suspension with DOCNa 20mM. The solid rhombs correspond to the surfactant concentrations selected to be used to diffusion measurements at thermodynamic equilibrium.*

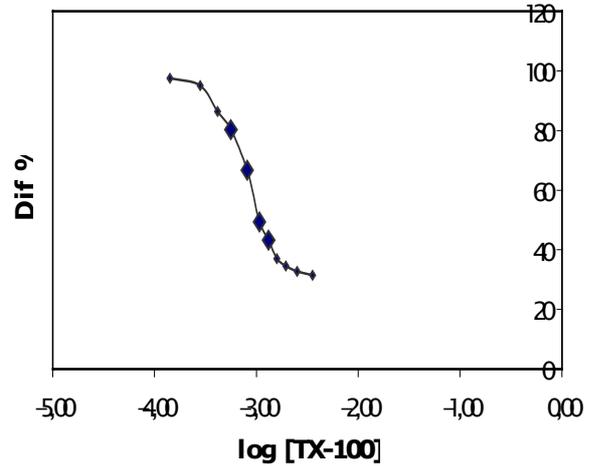

Figure 2 shows the exploring data (solid rhombus) of total concentration of DOCNa (Table 2) used for the determination of parameters $R_e^{sat}$ and $R_e^{sol}$ at thermodynamic equilibrium.

Table 3 presents the experimental data of titration with CTMB 20mM.

**Table 3.** *Rectangular optical diffusion (percentage), Dif%, of MLV-EYL in the course of titration with CTMB 20mM.*

| CTMB volume, µl | log[CTMB] | Dif% |
|---|---|---|
| 40 | -3.25 | 87.06 |
| 80 | -2.97 | 47.06 |
| 120 | -2.80 | 31.77 |
| 160 | -2.69 | 24.71 |
| 200 | -2.60 | 21.77 |
| 250 | -2.52 | 20.00 |
| 300 | -2.45 | 18.83 |
| 400 | -2.35 | 17.10 |
| 500 | -2.28 | 15.3 |
| 600 | -2.22 | 14.12 |
| 700 | -2.18 | 13.53 |
| 800 | -2.14 | 12.94 |

Figure 3 shows the exploring data (solid rhombus) of total concentration of CTMB (Table 3) used for the determination of parameters $R_e^{sat}$ and $R_e^{sol}$ at thermodynamic equilibrium.

**Figure 3.** *Titration curve of MLV-EYL suspension with CTMB 20mM. The solid rhombus correspond to the surfactant concentrations selected to be used to diffusion measurements at thermodynamic equilibrium.*

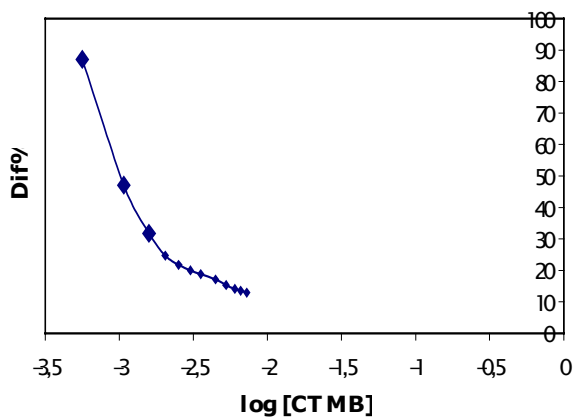

At thermodynamic equilibrium conditions were prepared the mixtures MLV-EYL-surfactant corresponding to the concentrations selected from the data of Figures 1, 2 and 3.

We worked with four series, consisting of 4-5 quots each, one of 1.4ml MLV-EYL suspension. The first series EYL had 0.25mM concentration, the second 0.50mM, the third 1.0mM and the fourth 1.43mM. In the samples of each series there were added surfactants so that the total concentrations ($D_t$) selected in the previous experiment should be obtained. After surfactant addition, each sample was vortexed and left at rest in order to take the equilibrium. After 16 hours the optical measurements at λ=437nm were made in order to obtain Dif% values, presented in Tables 4 - 7; 9 - 12 and 14 - 16.

**Table 4.** *Diffusimetric data at equilibrium in the case of TX-100. [EYL] =0.25mM.*

| TX-100 added volume, μl | log[TX-100] | Dif% |
|---|---|---|
| 40 | -3.25 | 68.75 |
| 60 | -3.09 | 42.50 |
| 80 | -2.97 | 42.50 |
| 100 | -2.88 | 33.75 |
| 1000 | -2.08 | - |

Figure 4 shows the experimental points which correlates the rectangular optic diffusions of MLV-surfactant mixtures ([EYL]=0.25mM) with total surfactant concentrations (Table 4). The $D_t^{sat}$ is the surfactant concentration which corresponds to intersection of regression line with the line of maximum value of optic diffusion.

**Figure 4.** *The graphic representation of Dif% dependence on log[TX-100]. [EYL] = 0.25mM. The interrupted straight line was obtained by least squares method (LS method).*

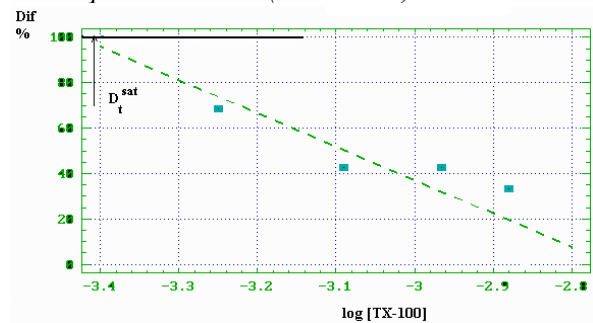

**Table 5.** *Diffusimetric data at equilibrium in the case of TX-100. [EYL ]=0.50mM.*

| TX-100 added volume, μliposomi | log[TX-100] | Dif% |
|---|---|---|
| 40 | -3.25 | 67.76 |
| 60 | -3.09 | 61.18 |
| 80 | -2.97 | 27.63 |
| 100 | -2.88 | 24.34 |
| 1000 | -2.08 | 21.05 |

Figure 5 shows the experimental points which correlates the rectangular optic diffusion of MLV-surfactant mixtures ([EYL]=0.50mM) with total surfactant concentration (Table 5). The $D_t^{sol}$ value is the surfactant concentration which corresponds to intersection of regression line with the line of optic difussion at the solubilization state.

**Figure 5.** *The graphic representation of Dif% dependence on log[TX-100]. [EYL]= 0.50mM. The interrupted straight line was obtained by least squares method (only square points).*

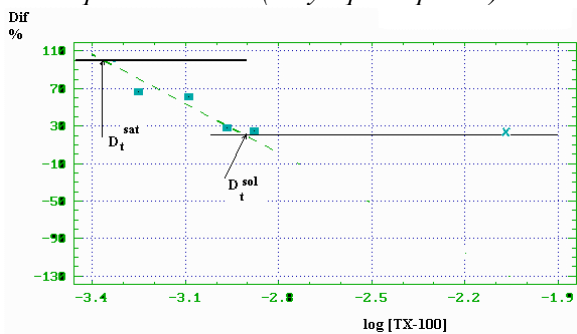

**Table 6.** *Diffusimetric data at equilibrium in the case of TX-100. [EYL]=1.0mM.*

| TX-100 added volume, μliposomi | log[TX-100] | Dif% |
|---|---|---|
| 40 | -3.25 | 80.11 |
| 60 | -3.09 | 65.90 |
| 80 | -2.97 | 47.72 |
| 100 | -2.88 | 43.75 |
| 1000 | -2.08 | 12.50 |

Figure 6 shows the experimental points which correlates the rectangular optic diffusions of MLV-surfactant mixtures ([EYL]=1.0mM) with total surfactant concentration (Table 6).

**Figure 6.** *The graphic representation of Dif% dependence on log[TX-100]. [EYL] = 1.0mM. The interrupted straight line was obtained by LS method (only square points).*

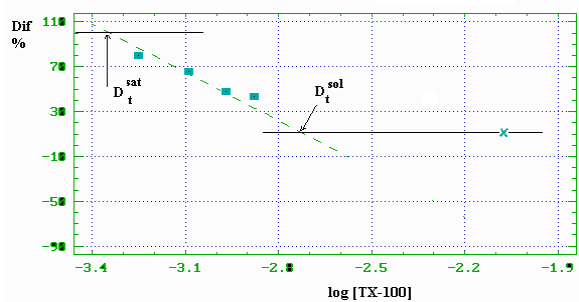

**Table 7.** *Diffusimetric data at equilibrium in the case of TX-100. [EYL]=1.43mM.*

| TX-100 added volume, μl | log[TX-100] | Dif% |
|---|---|---|
| 40 | -3.25 | 88.23 |
| 60 | -3.09 | 76.47 |
| 80 | -2.97 | 66.47 |
| 100 | -2.88 | 56.47 |
| 1000 | -2.08 | 9.99 |

Figure 7 shows the experimental points which correlates the rectangular optic diffusion of MLV-surfactant mixtures ([EYL]=1.43mM) with total surfactant concentration (Table 7).

**Figure 7.** *The graphic representation of Dif% dependence on log[TX-100]. [EYL] = 1.43mM. The interrupted straight line was obtained by LS method (only square points).*

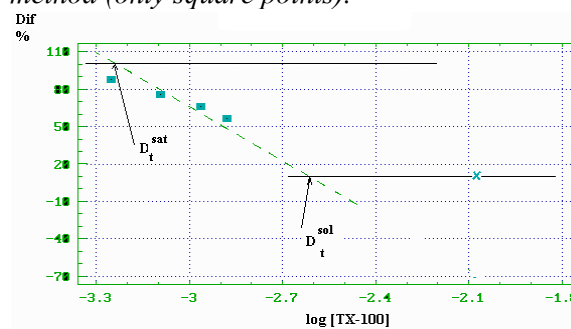

Summarizing of $D_t^{sat}$ and $D_t^{sol}$ values for TX-100 is in Table 8.

**Table 8.** *Synoptic tables of $D_t^{sat}$ and $D_t^{sol}$ values for TX-100.*

| MLV-EYL concn., mM | $D_t^{sat}$, mM | $D_t^{sol}$, mM |
|---|---|---|
| 0.25 | 0.390 | - |
| 0.50 | 0.436 | 1.26 |
| 1.00 | 0.447 | 1.86 |
| 1.43 | 0.575 | 2.42 |

The graphic representation of data in Table 8 appears in Figures 8 and 8A. The $D_t^{sat}$ and $D_t^{sol}$ straight line slopes are equal in number to $R_e^{sat}$ and $R_e^{sol}$ effective ratios. (Tab. 18).

**Figure 8.** *The graphic representation of $D_t^{sat}$, depending on EYL concentration, using TX-100 as surfactant. In this figure the $R_e^{sat}$ value was calculated from the $D_t^{sat}$ value straight line slope. The interrupted straight line was obtained by LS method.*

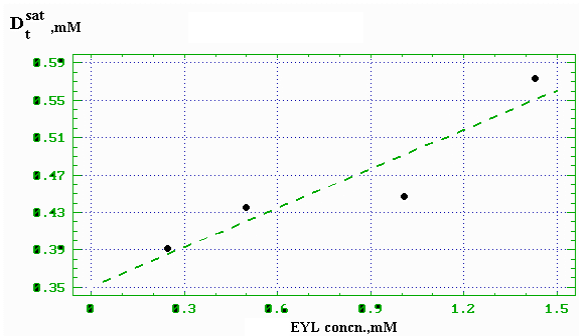

**Figure 8A.** *The graphic representation of $D_t^{sol}$, depending on EYL concentration, using TX-100 as surfactant. In this figure the $R_e^{sol}$ value was calculated from the $D_t^{sol}$ value straight line slope. The interrupted straight line was obtained by LS method.*

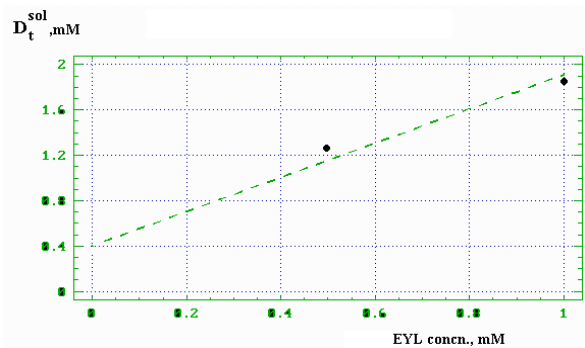

The literature reports [Partearroyo, 1992] these solubilization measurements with TX-100 for MLV and for LUV and for SUV-EYL. The obtained values in case MLV, where $R_e^{sat}$ =0.15 and $R_e^{sol}$=3.6 which differ a little from those we obtained. Our value, $R_e^{sat}$ = 0,14, is similar with the value reported in literature for MLV. It suggests that at saturation ($D_t^{sat}$, $R_e^{sat}$) a surfactant molecule is surrounded with seven phospholipid molecules. The fact that the $R_e^{sat}$ values obtained in literature (0.78 and 0.71 for SUV and LUV respectively) have much more higher values than that corresponding to MLV (0.15), is explained by the supposition that outer bilayers of MLV are solubilized, through a "peeling off" process before the system reaches the equilibrium. This process determines a solubilization with a biphasic kinetics: the solubilization of the outer bilayers takes place in seconds-tenth seconds (slow phase); the equilibration of the remnant vesicles with water environment takes place in several hours (very slow phase). In our experiment at the concentration at which the vesicles are transformed into mixed micelles, the 1.5 value of solubilising effective ratio, $R_e^{sol}$, as it was expected, has a much higher value than $R_e^{sat}$. It suggest that, in a mixed micelle, ten lipid molecules coexist with about fifteen surfactant molecules. Between $R_e^{sat}$ and $R_e^{sol}$ values there are coexisting phospholipid-surfactant mixed vesicles with phospholipid-surfactant mixed micelles. This interval is called "coexisting interval".

In the case of interaction of liposomes with TX-100 we have $D_w^{sat}= D_w^{sol}= D_w^0$. The $D_w^0$ value is approximate with critical micellar concentration (CMC) value. Figure 8 shows $D_w^0$ =0.34 mM, a value higher than CMC reported in literature (Table 19).

**Table 9.** *Diffusimetric data at equilibrium in the case of DOCNa. [EYL]=0.25mM.*

| DOCNa added volume, μl | log[DOCNa] | Dif% |
|---|---|---|
| 80 | -2.97 | 64.7 |
| 160 | -2.69 | 58.8 |
| 200 | -2.60 | 38.2 |
| 1000 | -2.08 | 27.5 |

Figure 9 shows the experimental points which correlates the rectangular optic diffusion of MLV-surfactant mixtures ([EYL]=0.25mM) with total surfactant concentration (Table 9).

**Figure 9.** *The graphic representation of Dif% dependence on log[DOCNa]. [EYL] = 0.25mM. The interrupted straight line was obtained by LS method (only square points).*

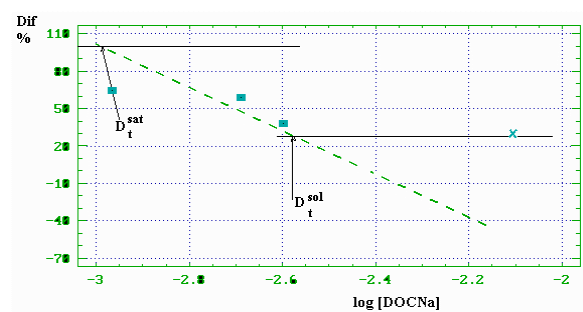

**Table 10.** *Diffusimetric data at equilibrium in the case of DOCNa. [EYL]=0.50mM.*

| DOCNa added volume, μl | log[DOCNa] | Dif% |
|---|---|---|
| 80 | -2.97 | 69.0 |
| 120 | -2.80 | 51.2 |
| 160 | -2.69 | 49.4 |
| 200 | -2.60 | 38.7 |
| 1000 | -2.08 | 23.8 |

Figure 10 shows the experimental points which correlates the rectangular optic diffusion of MLV-surfactant mixtures ([EYL]=0.50mM) with total surfactant concentration (Table 10).

**Figure 10.** *The graphic representation of Dif% dependence on log[DOCNa]. [EYL] = 0.50mM. The interrupted straight line was obtained by LS method (only square points).*

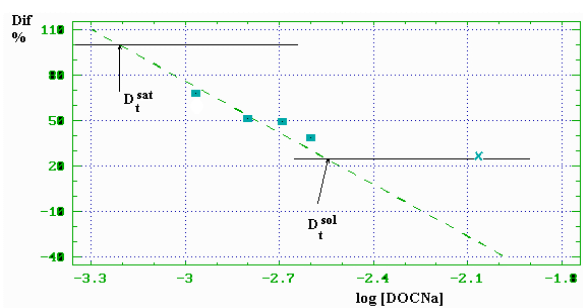

**Table 11.** *Diffusimetric data at equilibrium in the case of DOCNa. [EYL]=1.0mM.*

| DOCNa added volume, μl | log[DOCNa] | Dif% |
|---|---|---|
| 80 | -2.97 | 69.9 |
| 120 | -2.80 | 53.4 |
| 160 | -2.69 | 51.1 |
| 200 | -2.60 | 40.9 |
| 1000 | -2.08 | 14.8 |

Figure 11 shows the experimental points which correlates the rectangular optic diffusion of MLV-surfactant mixtures ([EYL]=1.0mM with total surfactant concentration (Table 11).

**Figure 11.** *The graphic representation of Dif% dependence on log[DOCNa]. [EYL] = 1.0mM. The interrupted straight line was obtained by LS method (only square points).*

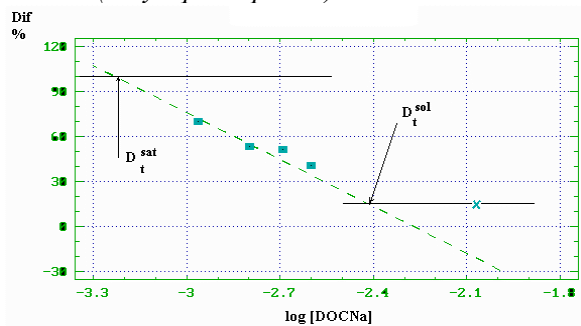

**Table 12.** *Diffusimetric data at equilibrium in the case of DOCNa. [EYL]=1.43mM.*

| DOCNa added volume, μl | log[DOCNa] | Dif% |
|---|---|---|
| 80 | -2.97 | 64.4 |
| 120 | -2.80 | 55.2 |
| 160 | -2.69 | 41.4 |
| 200 | -2.60 | 35.6 |
| 1000 | -2.08 | 14.4 |

Figure 12 shows the experimental points which correlates the rectangular optic diffusion of MLV-surfactant mixtures ([EYL] =1.43mM) with total surfactant concentration (Table 12).

**Figure 12.** *The graphic representation of Dif% dependence on log[DOCNa]. [EYL] = 1.43mM. The interrupted straight line was obtained by LS method (only square points).*

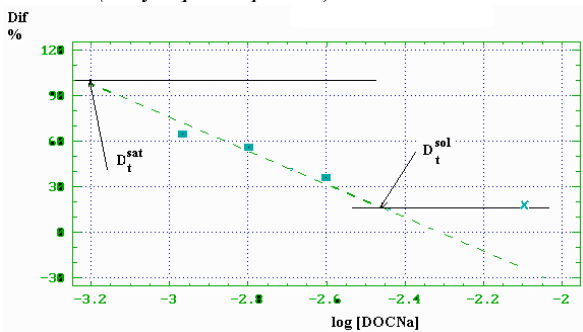

**Table 13.** *Synoptic table of $D_t^{sat}$ and $D_t^{sol}$ values for DOCNa.*

| MLV-EYL concn., mM | $D_t^{sat}$, mM | $D_t^{sol}$, mM |
|---|---|---|
| 0.25 | 1.06 | 2.66 |
| 0.50 | 0.616 | 2.82 |
| 1.0 | 0.600 | 3.85 |
| 1.43 | 0.630 | 3.47 |

The graphic representation of data in Table 13 appears in Figures 13 and 13A. The $R_e^{sat}$ value is very close by approaching with effective ratio value corresponding to the beginning of solubilization (=0.1) obtained in case of MLV-EYL sodium with taurodeoxycholate interaction, the study made through rapid ultrafiltration [Müller, 1990]. The values of $R_e^{sat}$ suggest that MLV saturate with DOCNa when in the membrane one surfactant molecule for 14 phospholipid molecules each is present. The solubilization is finished when a surfactant molecule corresponds approximately with a phospholipid molecule, this being the numeric ratio between the molecules of the two species present in the mixed micelles at the end of the mixed vesicles→mixed micelles transformation process.

**Figure 13.** *The graphic representation of $D_t^{sat}$ dependence on EYL concentration using DOCNa as surfactant. The $R_e^{sat}$ value was calculated from the $D_t^{sat}$ straight line slope. The interrupted straight line was obtained by LS method.*

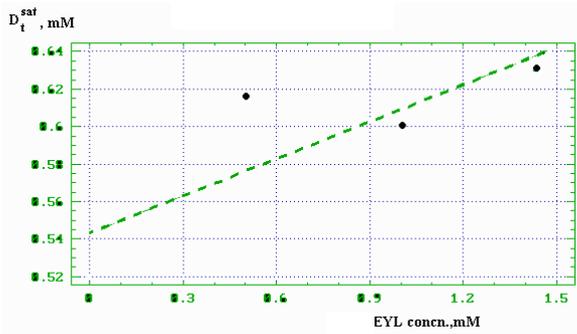

**Figure 13A.** *The graphic representation of $D_t^{sol}$ dependence on EYL concentration using DOCNa as surfactant. The $R_e^{sol}$ value was calculated from the $D_t^{sol}$ straight line slope. The interrupted straight line was obtained by LS method.*

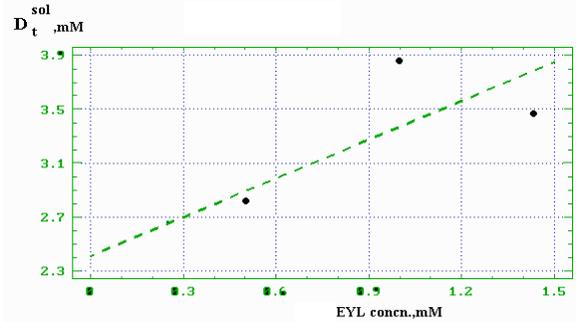

**Table 14.** *Diffusimetric data at equilibrium in the case of CTMB. [EYL]=0.25mM.*

| CTMB added volume, μl | log[CTMB] | Dif% |
|---|---|---|
| 40 | -3.25 | 36.3 |
| 60 | -3.08 | 21.6 |
| 80 | -2.97 | 18.6 |
| 120 | -2.80 | 17.6 |
| 1000 | -2.08 | 15.7 |

Figure 14 shows the experimental points which correlates the rectangular optic diffusion of MLV-surfactant mixtures ([EYL]=0.25mM with total surfactant concentration (Table 14).

**Figure 14.** *The graphic representation of Dif% dependence on log[CTMB]. [EYL]= 0.25mM. The interrupted straight line was obtained by LS method (only square points).*

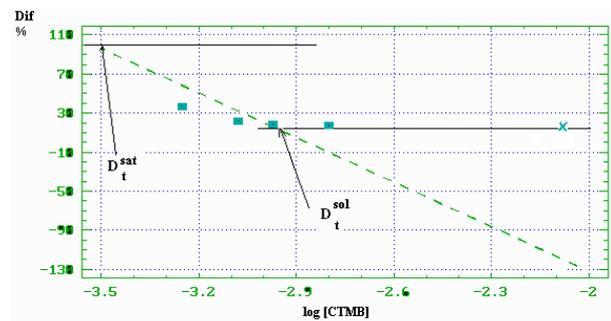

**Table 15.** *Diffusimetric data at equilibrium in the case of CTMB. [EYL]=0.50mM.*

| CTMB added volume, μl | log[CTMB] | Dif% |
|---|---|---|
| 40 | -3.25 | - |
| 60 | -3.08 | 42.8 |
| 80 | -2.97 | 34.5 |
| 120 | -2.80 | 16.7 |
| 1000 | -2.08 | 22.6 |

Figure 15 shows the experimental points which correlates the rectangular optic diffusion of MLV-surfactant mixtures ([EYL]=0.50mM) with total surfactant concentration (Table 15).

**Figure 15.** *The graphic representation of Dif% dependence on log[CTMB]. [EYL] = 0.50mM. The interrupted straight line was obtained by LS method (only square points).*

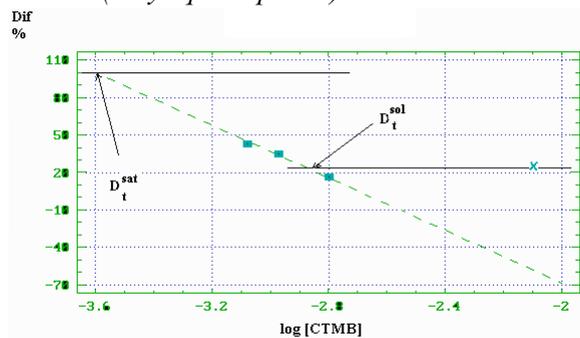

**Table 16.** *Diffusimetric data at equilibrium in the case of CTMB. [EYL]=1.43mM.*

| CTMB added volume, μl | log[CTMB] | Dif% |
|---|---|---|
| 40 | -3.25 | - |
| 60 | -3.08 | 90.0 |
| 80 | -2.97 | 64.8 |
| 120 | -2.80 | 60.5 |
| 1000 | -2.08 | 1.20 |

Figure 16 shows the experimental points which correlates the rectangular optic diffusion of MLV-surfactant mixtures ([EYL]=1.43mM) with total surfactant concentration (Table 16).

**Figure 16.** *The graphic representation of Dif% dependence on log[CTMB]. [EYL] = 1.43mM. The interrupted straight line was obtained by LS method (only square points).*

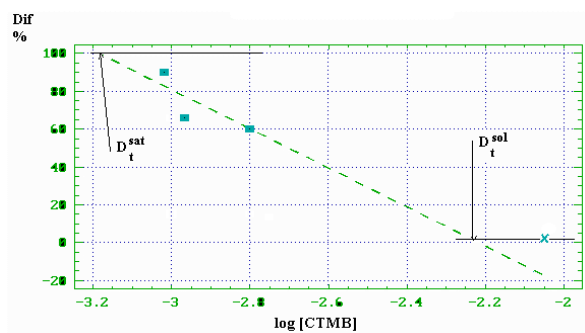

**Table 17.** *Synoptic table of $D_t^{sat}$ and $D_t^{sol}$ values for CTMB.*

| MLV-EYL concn., mM | $D_t^{sat}$, mM | $D_t^{sol}$, mM |
|---|---|---|
| 0.25 | 0.32 | 1.12 |
| 0.50 | 0.26 | 1.38 |
| 1.43 | 0.66 | 5.96 |

The graphic representation of data in Table 17 appears in Figures 17 and 17A.

The $R_e^{sat}$ and $R_e^{sol}$ values indicate the fact that, at the onset of the conversion process *mixed vesicles→mixed micelles*, in bilayers the EYL and CTMB molecules are in the molecular ratio 4:1 and at the end of the solubilization process in the mixed micelles the phospholipid molecules and surfactant molecules are in the molecular ratio 1:4.

**Figure 17.** *The graphic representation of $D_t^{sol}$ dependence on EYL concentration, using CTMB as surfactant. The $R_e^{sol}$ value was calculated from the $D_t^{sol}$ straight line slope (LS method).*

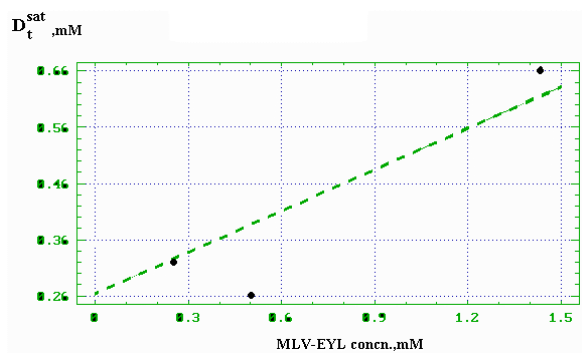

**Figure 17A.** *The graphic representation of $D_t^{sol}$ dependence on EYL concentration, using CTMB as surfactant. The $R_e^{sol}$ value was calculated from the $D_t^{sol}$ straight line slope (LS method).*

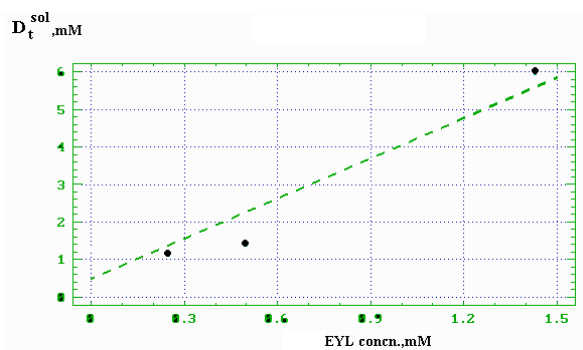

**Table 18.** *Synoptic table of $R_e^{sat}$ and $R_e^{sol}$ values for studied surfactants.*

| Surfactant | $R_e^{sat}$ | $R_e^{sol}$ |
|---|---|---|
| DOCNa | 0.07 | 1.1 |
| Triton X-100 | 0.14 | 1.5 |
| CTMB | 0.25 | 3.6 |

Table 18 shows that the $R_e$ (*sat* and *sol*) value sequence for the three surfactant is $R_e^{DOCNa} < R_e^{TX-100} < R_e^{CTMB}$. The solubilization power thermodynamic equilibrum for the three surfactants is in the sequence: CTMB < TX-100 < DOCNa.

Table 19 is summary of the CMC values for the studied surfactants, reported in literature and those experimentally obtained by us.

**Table 19.** *The critical micellar concentration (CMC) values for the studied surfactants are reported both in literature and by our experiments.*

| Surfactant | CMC in literature (mM) | CMC (mM) |
|---|---|---|
| DOCNa | 1-5 | 0.54-2.35 |
| Triton X-100 | 0.2-0.9 | 0.34 |
| CTMB | 0.8-1.0 | 0.25 |

A greater difference is observed in CTMB case, due to water environment composition. In our experiments buffer Tris-Cl 0.05M, pH=7.2 was used.

**Figure 1.** *Titration curve of MLV-EYL suspension with TX-100. The solid rhombs correspond to the surfactant concentrations selected to be used to diffusion measurements at thermodynamic equilibrium.*

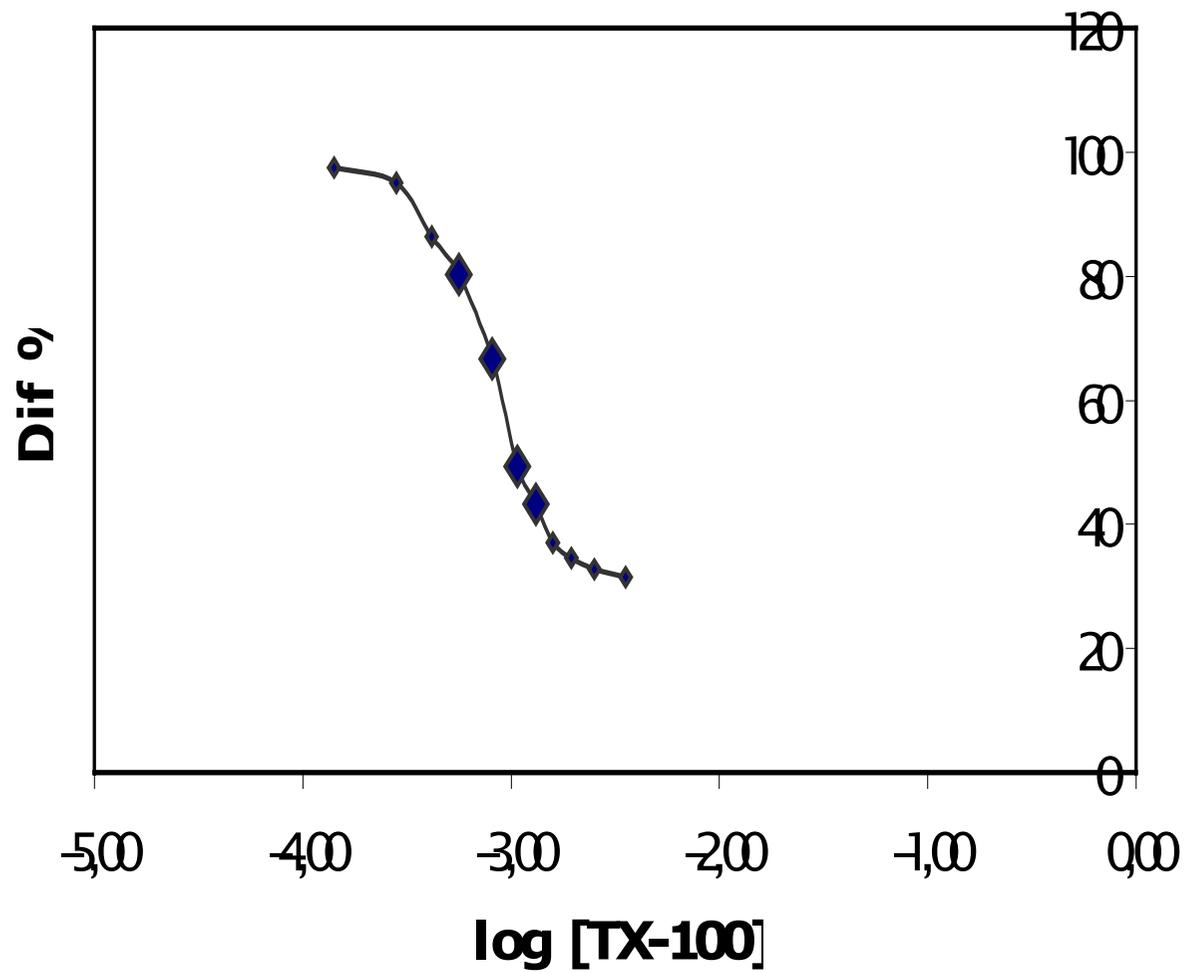

**Figure 2.** *Titration curve of MLV-EYL suspension with DOCNa 20mM. The solid rhombs correspond to the surfactant concentrations selected to be used to diffusion measurements at thermodynamic equilibrium.*

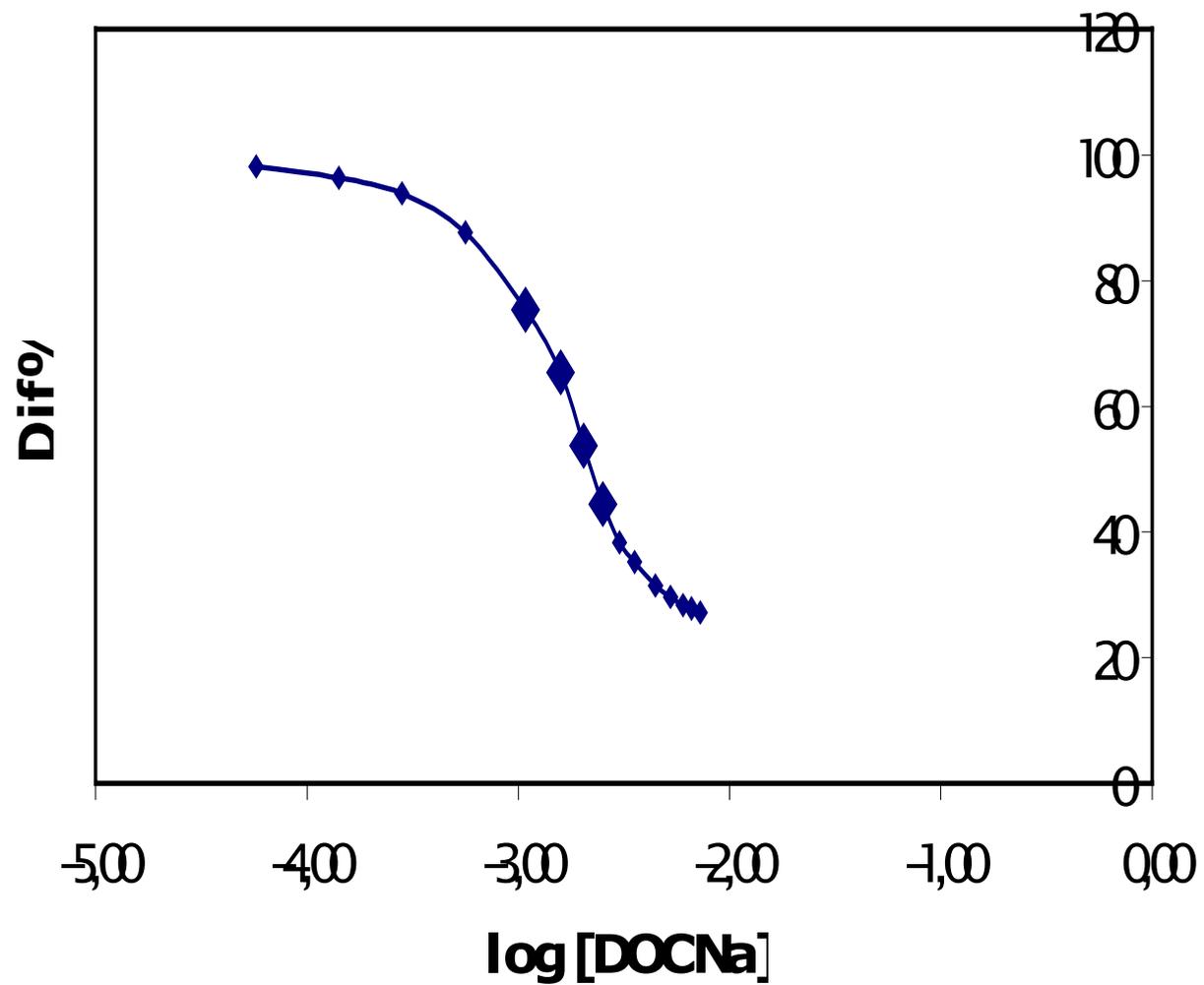

**Figure 3.** *Titration curve of MLV-EYL suspension with CTMB 20mM. The solid rhombus correspond to the surfactant concentrations selected to be used to diffusion measurements at thermodynamic equilibrium.*

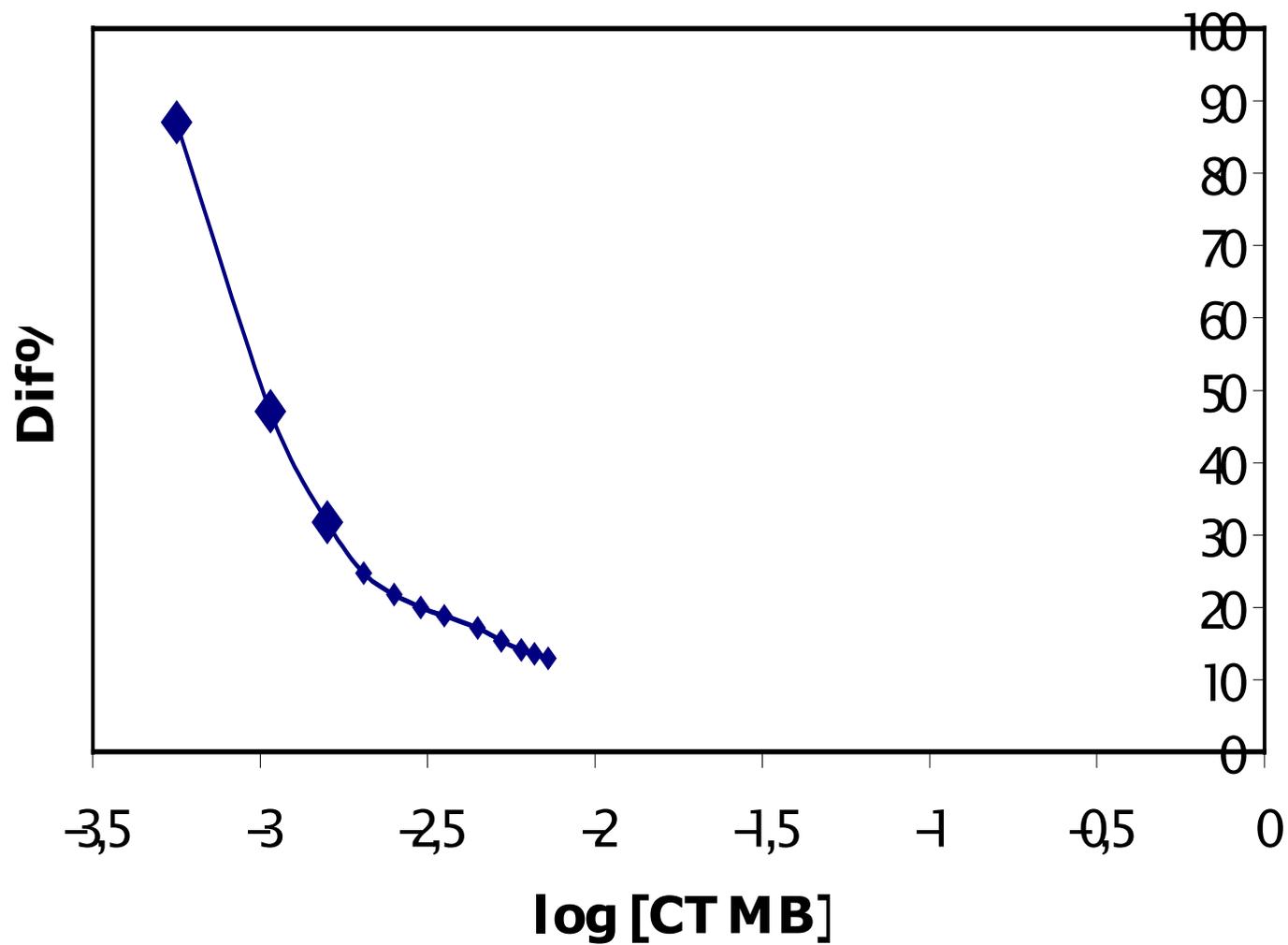

**Figure 4.** *The graphic representation of Dif% dependence on log[TX-100]. [EYL] = 0.25mM. The interrupted straight line was obtained by least squares method (LS method).*

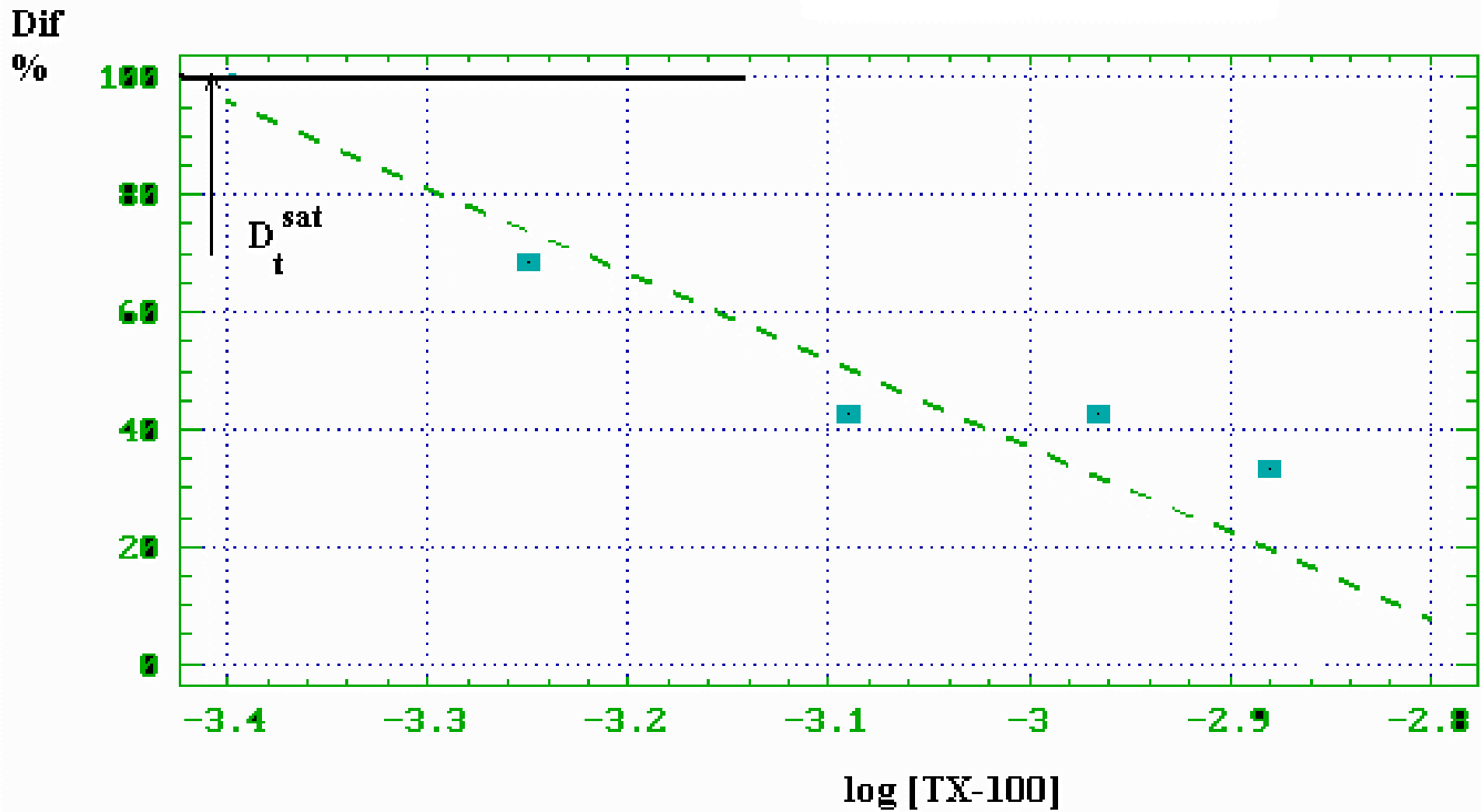

**Figure 5.** *The graphic representation of Dif% dependence on log[TX-100]. [EYL]= 0.50mM. The interrupted straight line was obtained by least squares method (only square points).*

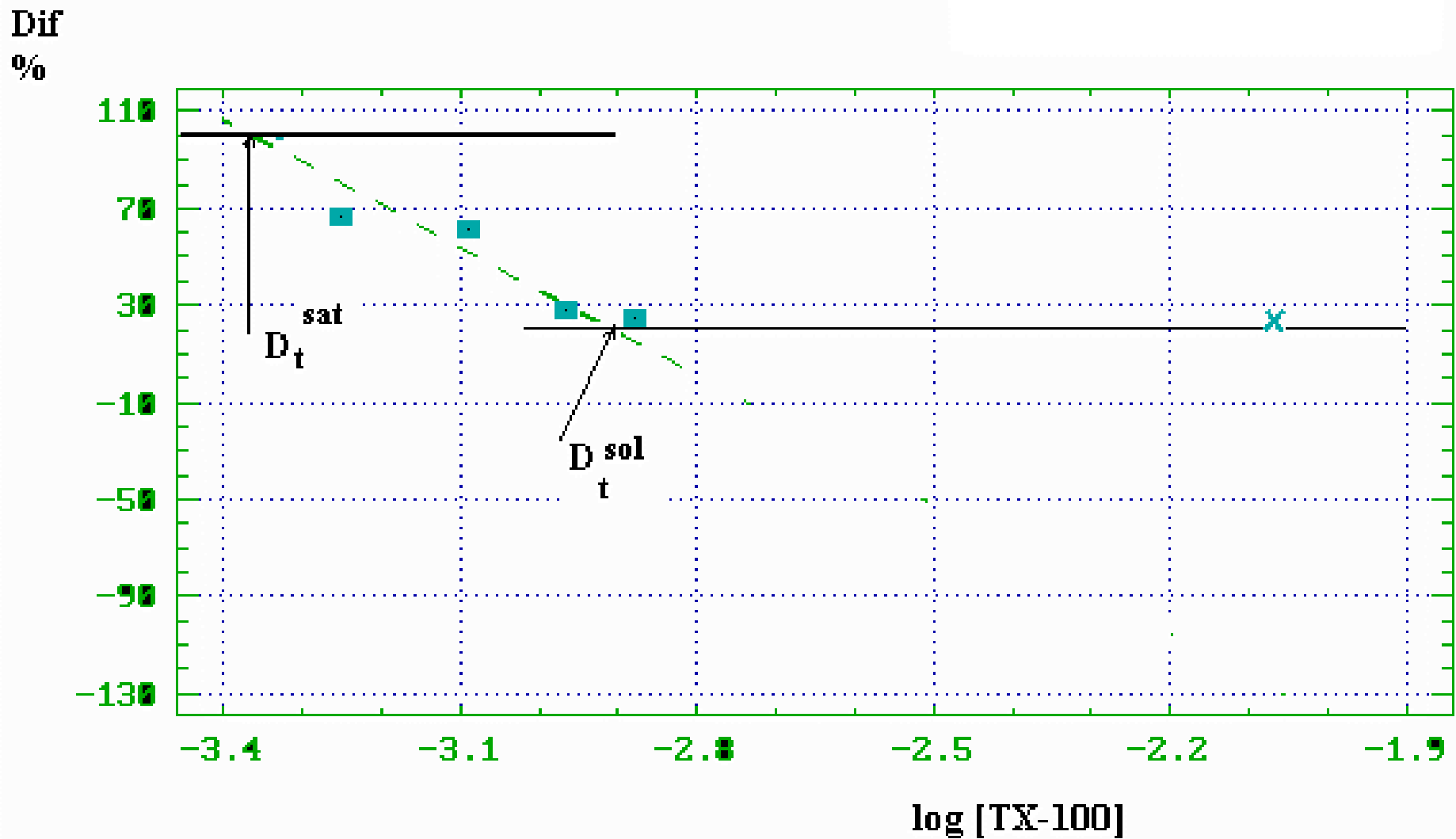

**Figure 6.** *The graphic representation of Dif% dependence on log[TX-100]. [EYL] = 1.0mM. The interrupted straight line was obtained by LS method (only square points).*

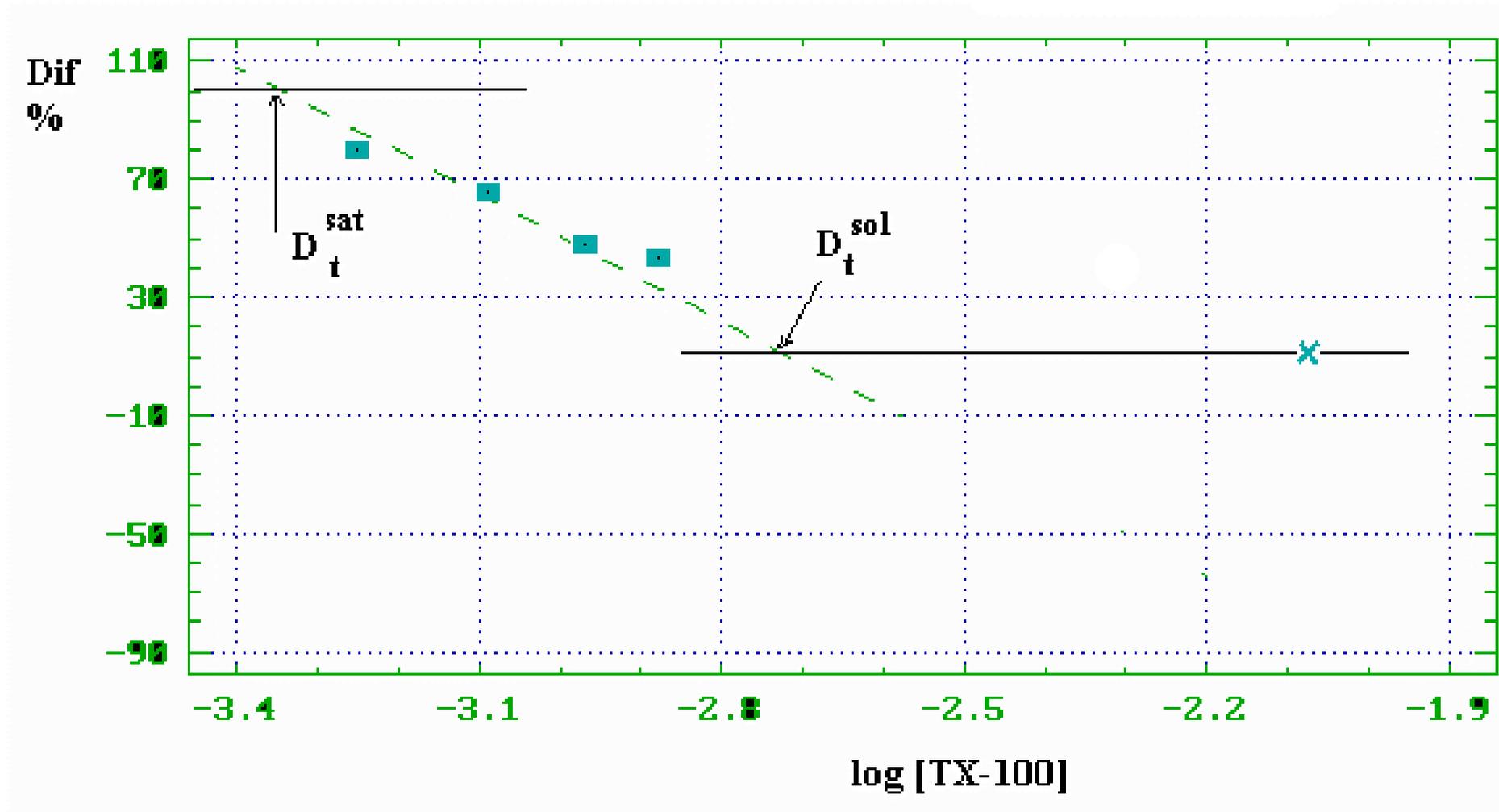

**Figure 7.** *The graphic representation of Dif% dependence on log[TX-100]. [EYL] = 1.43mM. The interrupted straight line was obtained by LS method (only square points).*

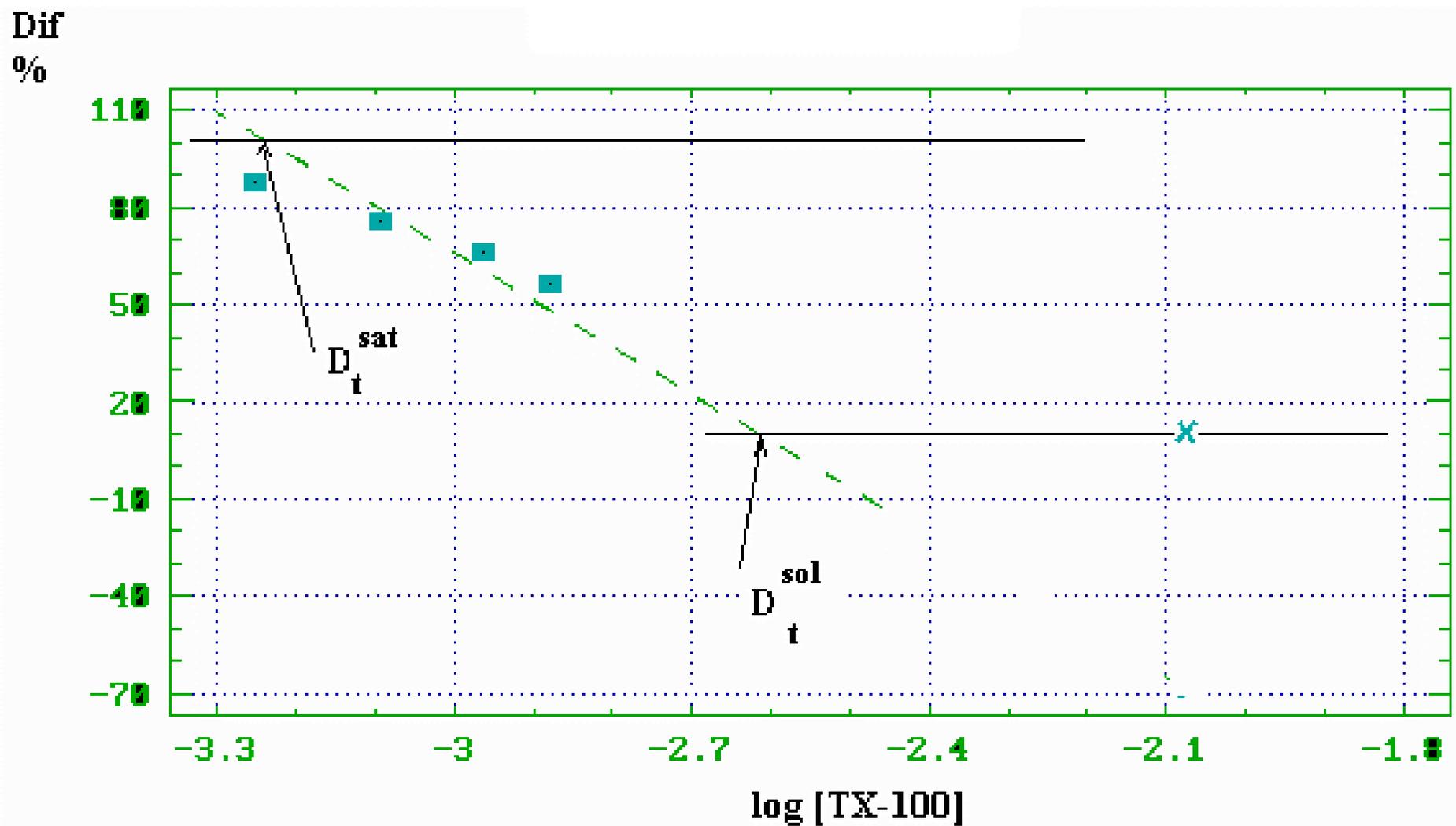

**Figure 8.** *The graphic representation of $D_t^{sat}$, depending on EYL concentration, using TX-100 as surfactant. In this figure the $R_e^{sat}$ value was calculated from the $D_t^{sat}$ value straight line slope.*

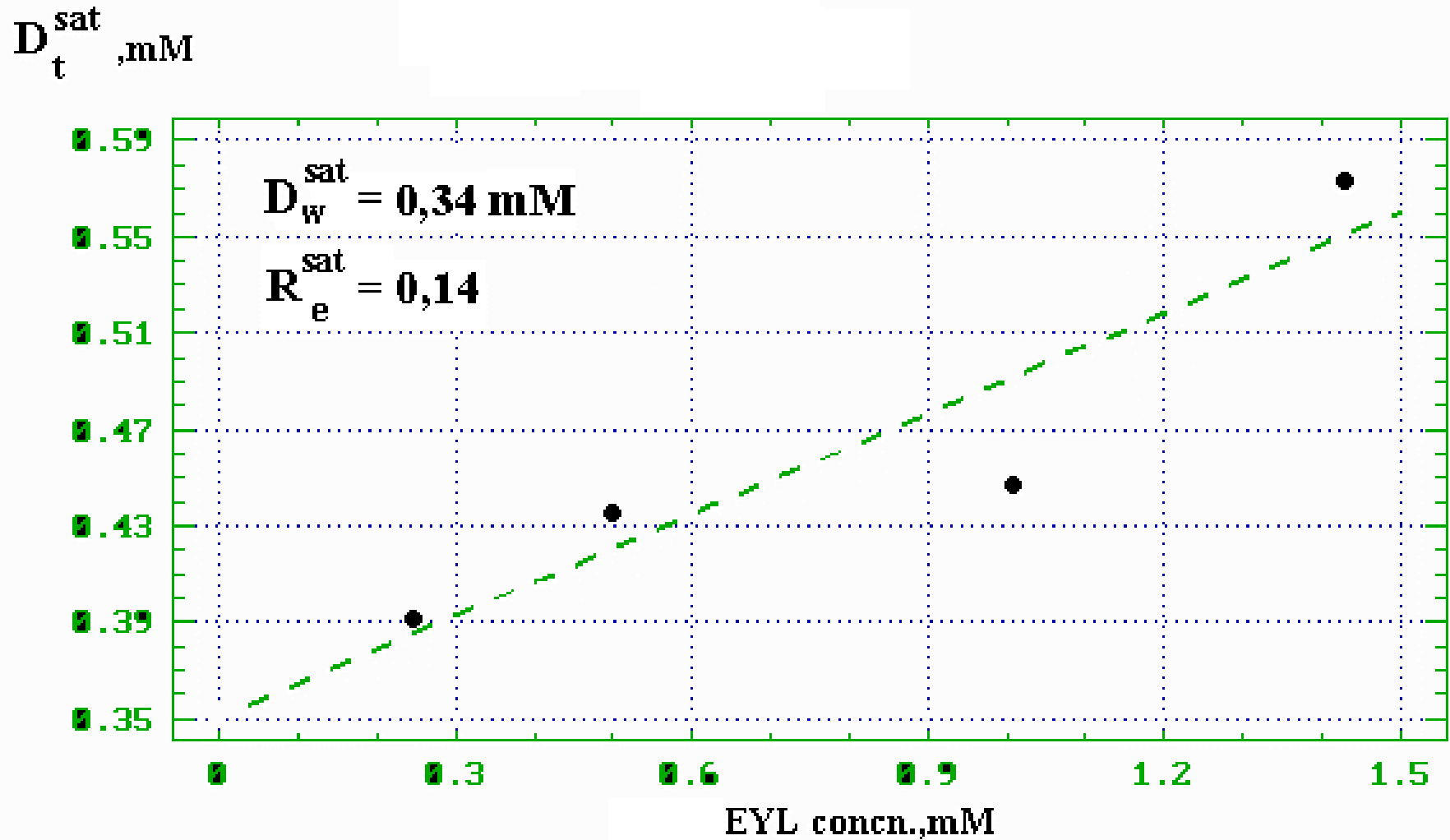

**Figure 8A.** *The graphic representation of $D_t^{sol}$, depending on EYL concentration, using TX-100 as surfactant. In this figure the $R_e^{sol}$ value was calculated from the $D_t^{sol}$ value straight line slope.*

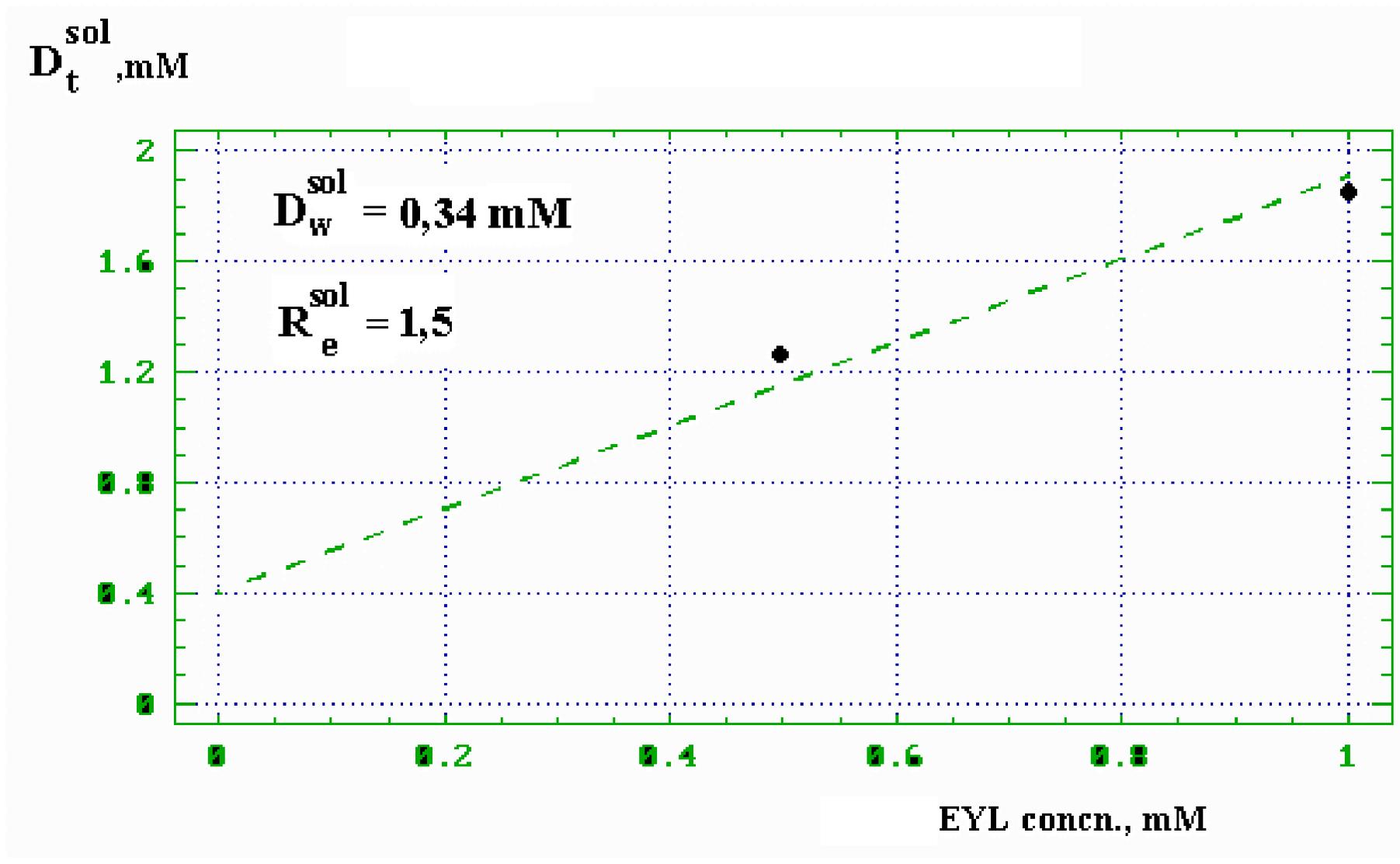

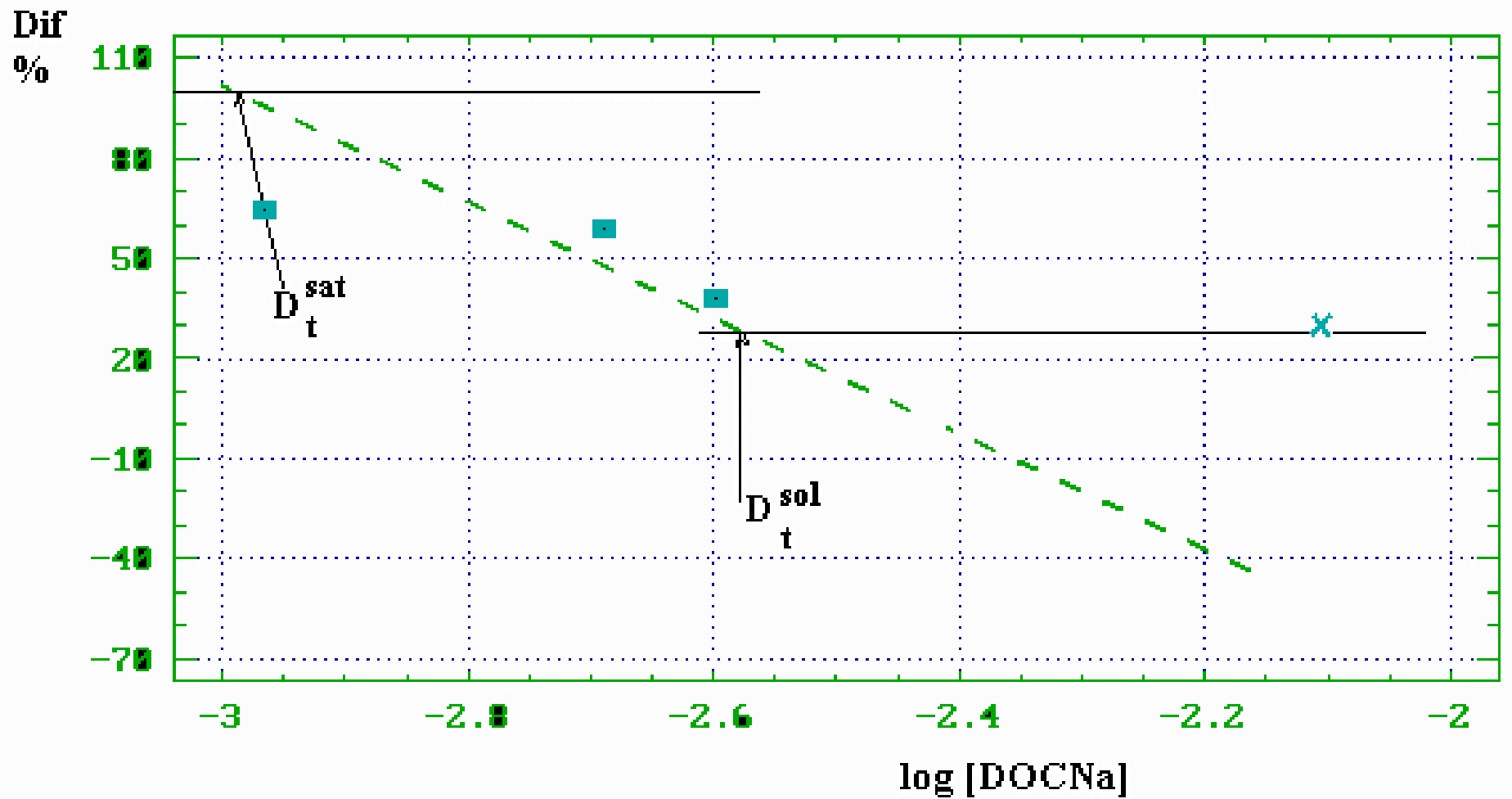

**Figure 9.** *The graphic representation of Dif% dependence on log[DOCNa]. [EYL] = 0.25mM. The interrupted straight line was obtained by LS method (only square points).*

**Figure 10.** *The graphic representation of Dif% dependence on log[DOCNa]. [EYL] = 0.50mM. The interrupted straight line was obtained by LS method (only square points).*

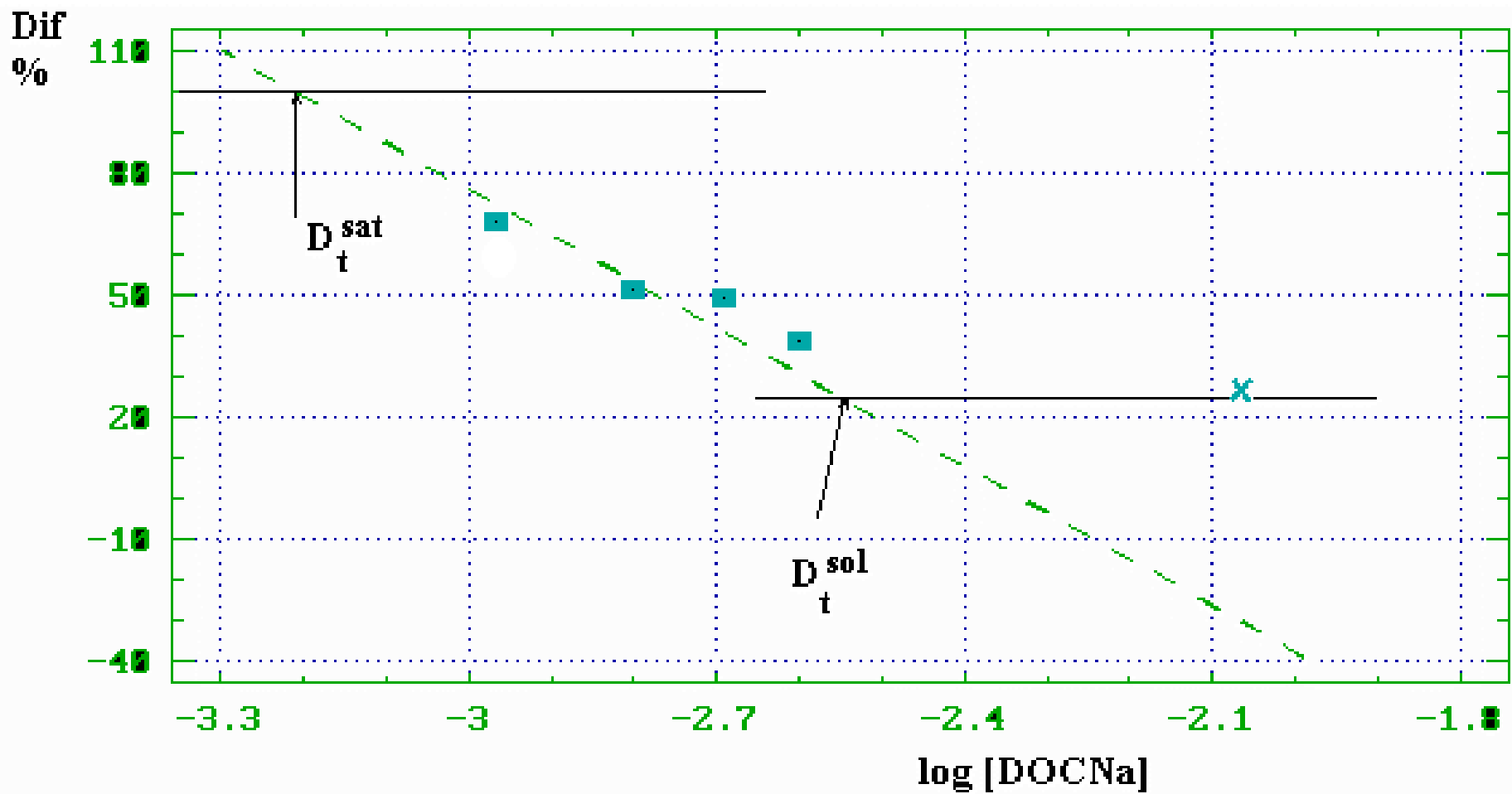

**Figure 11.** *The graphic representation of Dif% dependence on log[DOCNa]. [EYL] = 1.0mM. The interrupted straight line was obtained by LS method (only square points).*

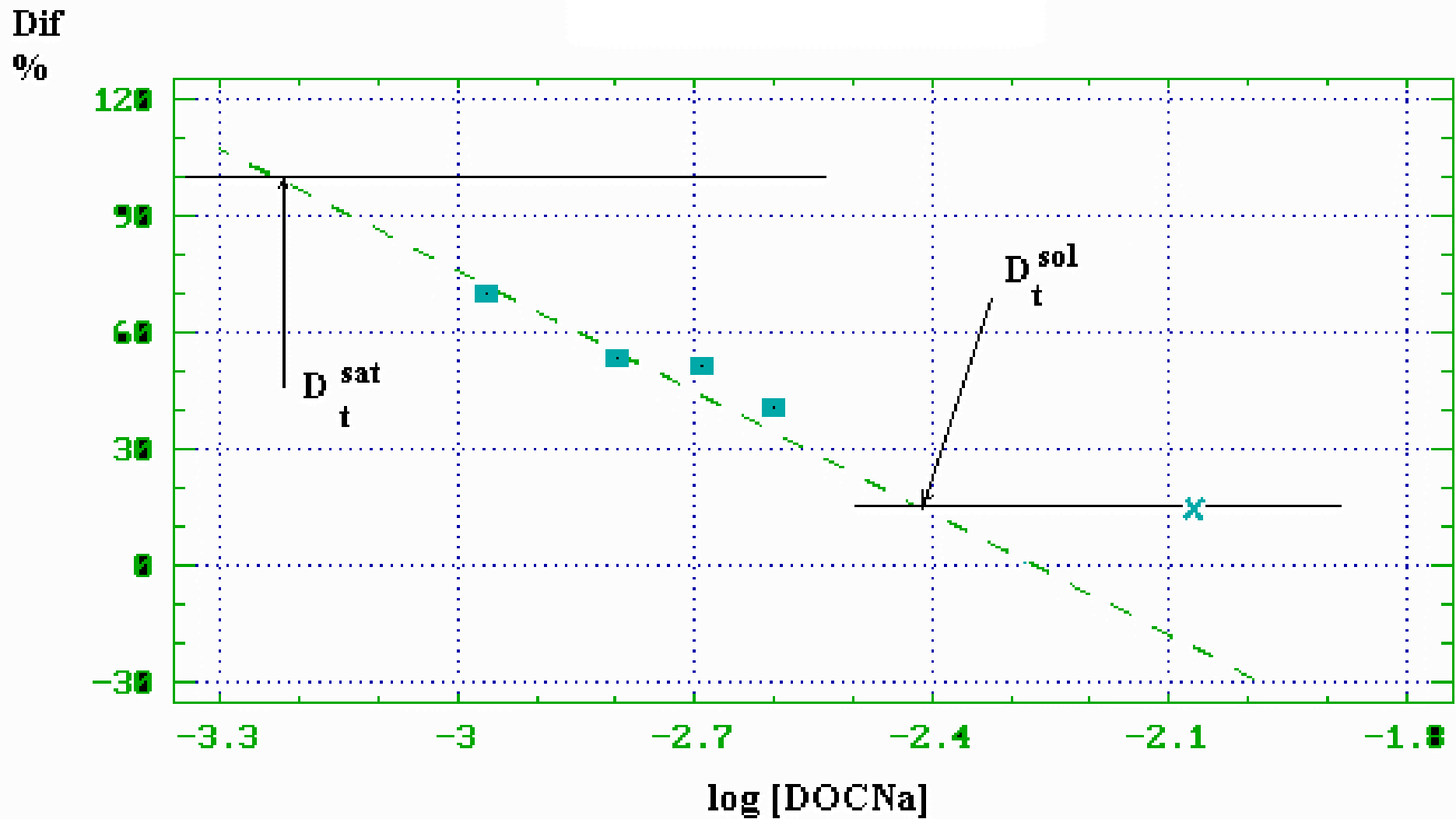

**Figure 12.** *The graphic representation of Dif% dependence on log[DOCNa]. [EYL] = 1.43mM. The interrupted straight line was obtained by LS method (only square points).*

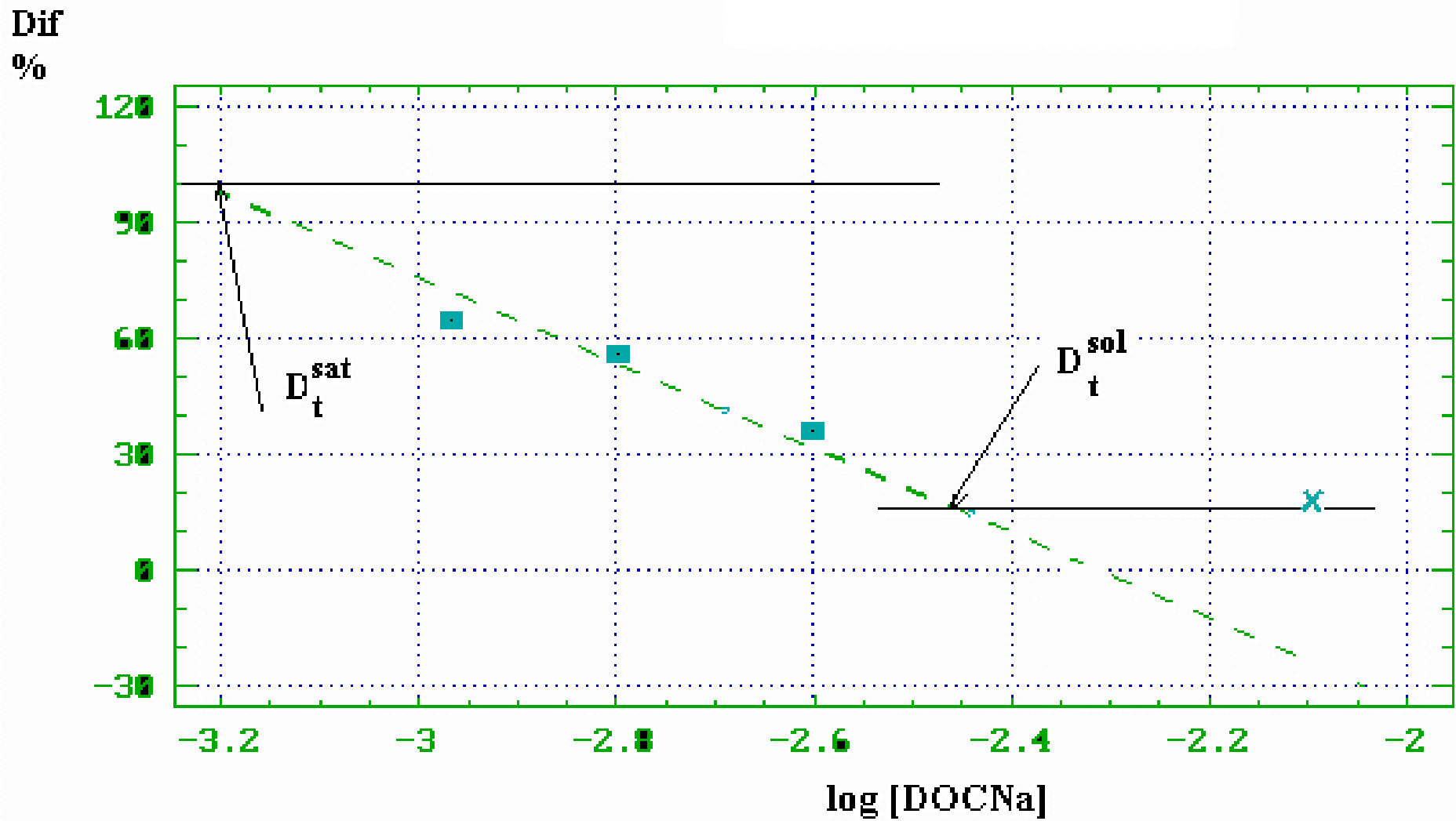

**Figure 13.** *The graphic representation of $D_t^{sat}$ dependence on EYL concentration using DOCNa as surfactant. The $R_e^{sat}$ value was calculated from the $D_t^{sat}$ straight line slope.*

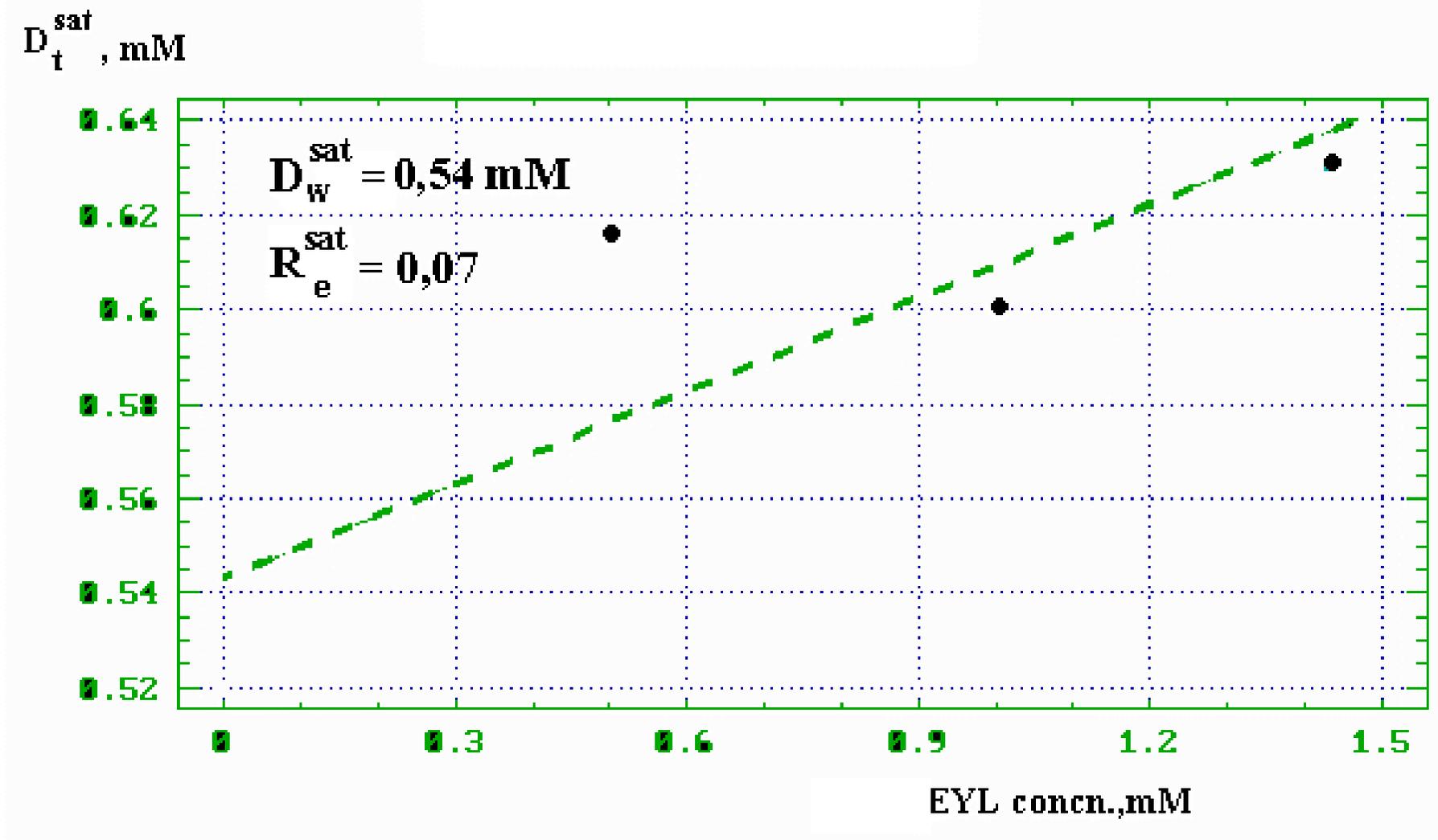

**Figure 13A.** *The graphic representation of $D_t^{sol}$ dependence on EYL concentration using DOCNa as surfactant. The $R_e^{sol}$ value was calculated from the $D_t^{sol}$ straight line slope.*

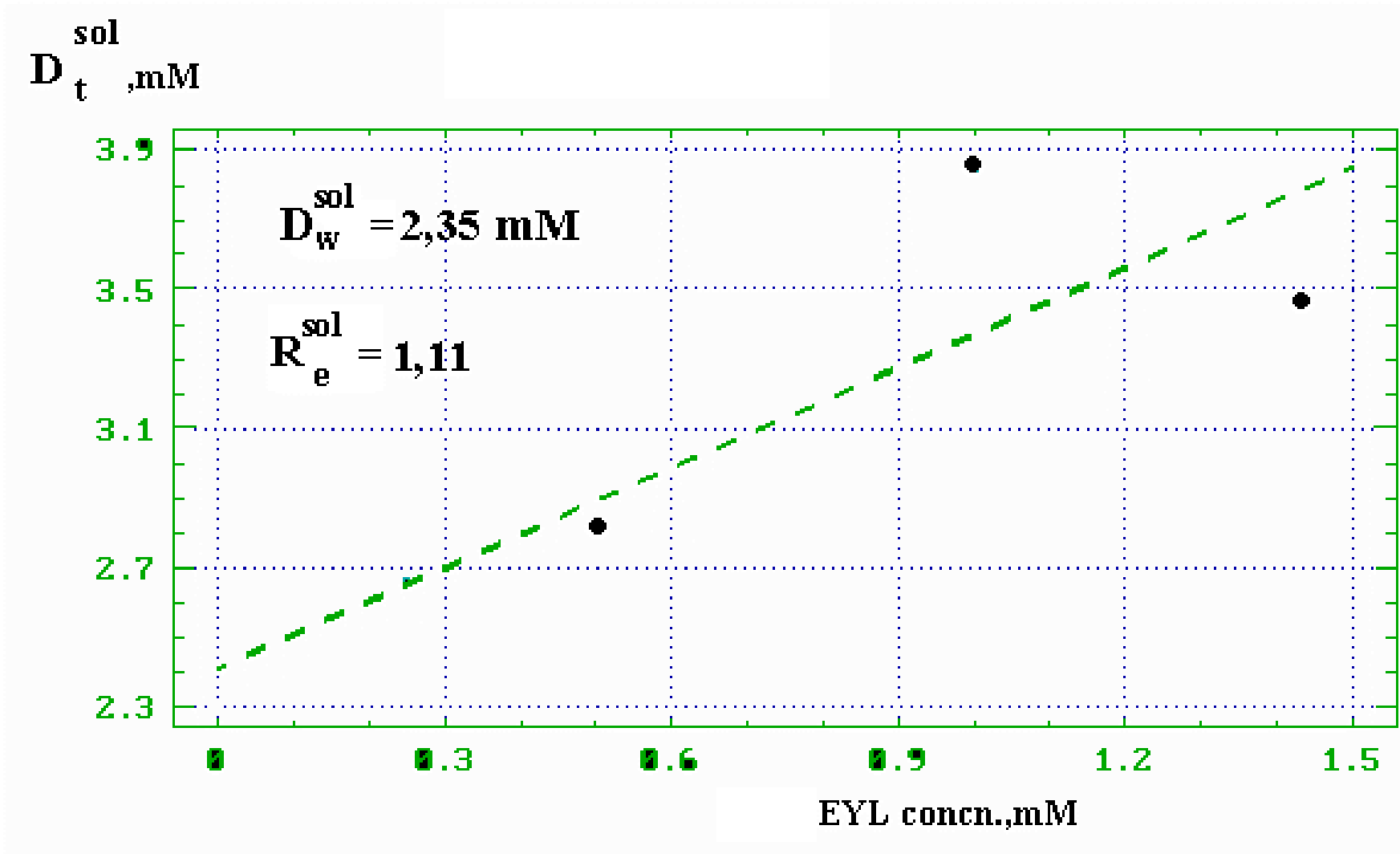

**Figure 14.** *The graphic representation of Dif% dependence on log[CTMB]. [EYL]= 0.25mM. The interrupted straight line was obtained by LS method (only square points).*

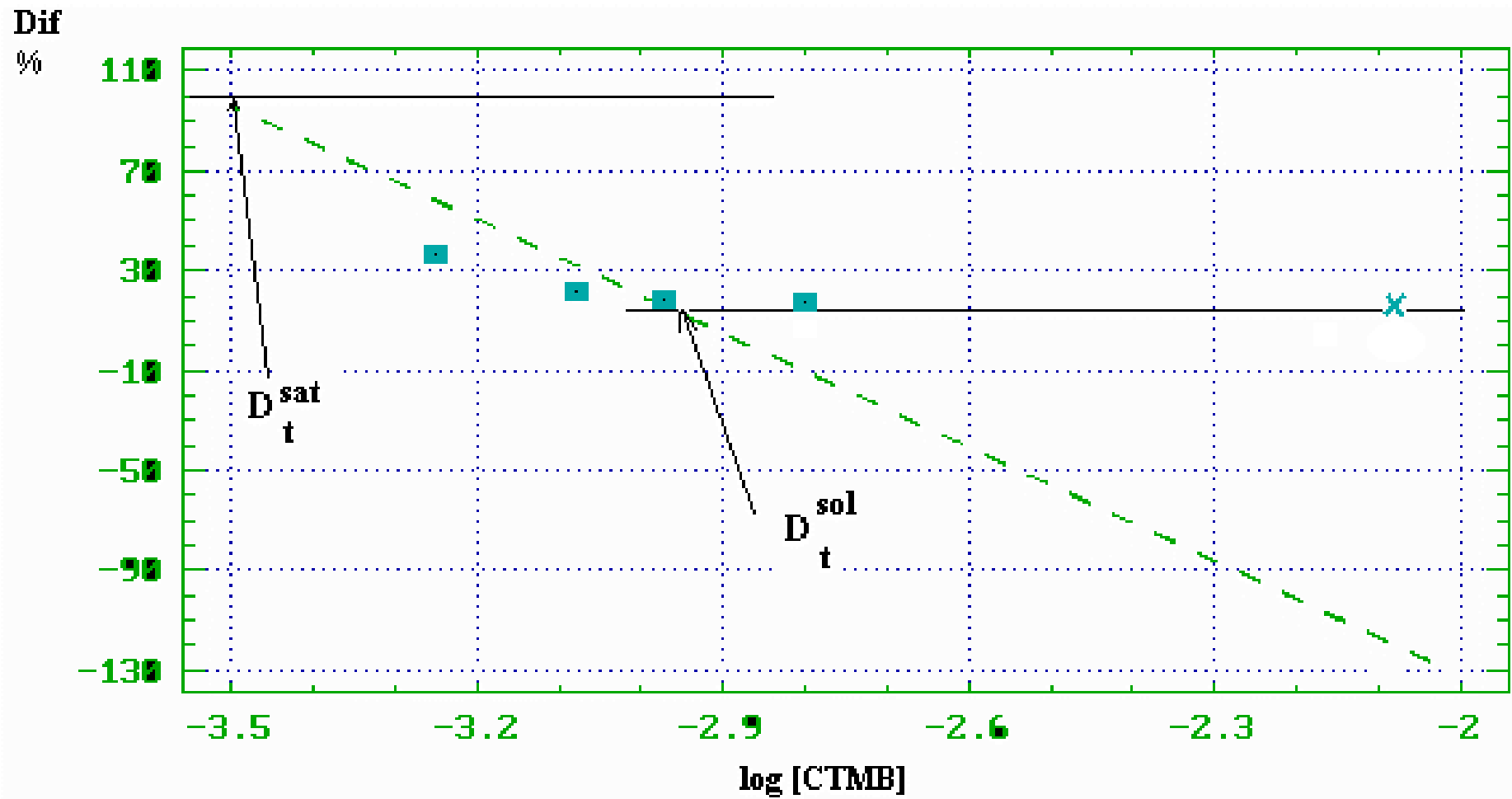

**Figure 15.** *The graphic representation of Dif% dependence on log[CTMB]. [EYL] = 0.50mM. The interrupted straight line was obtained by LS method (only square points).*

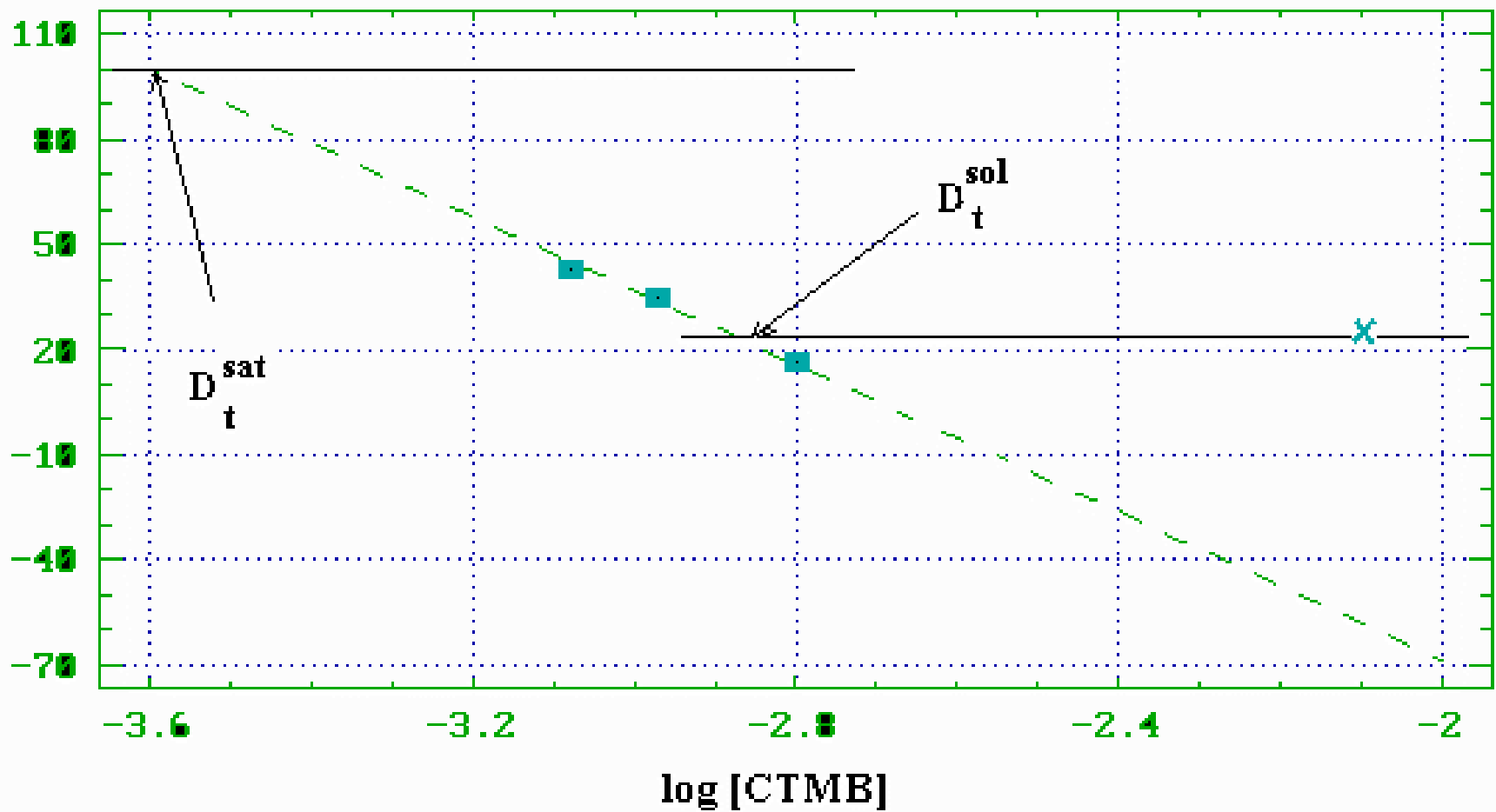

**Figure 16.** *The graphic representation of Dif% dependence on log[CTMB]. [EYL] = 1.43mM. The interrupted straight line was obtained by LS method (only square points).*

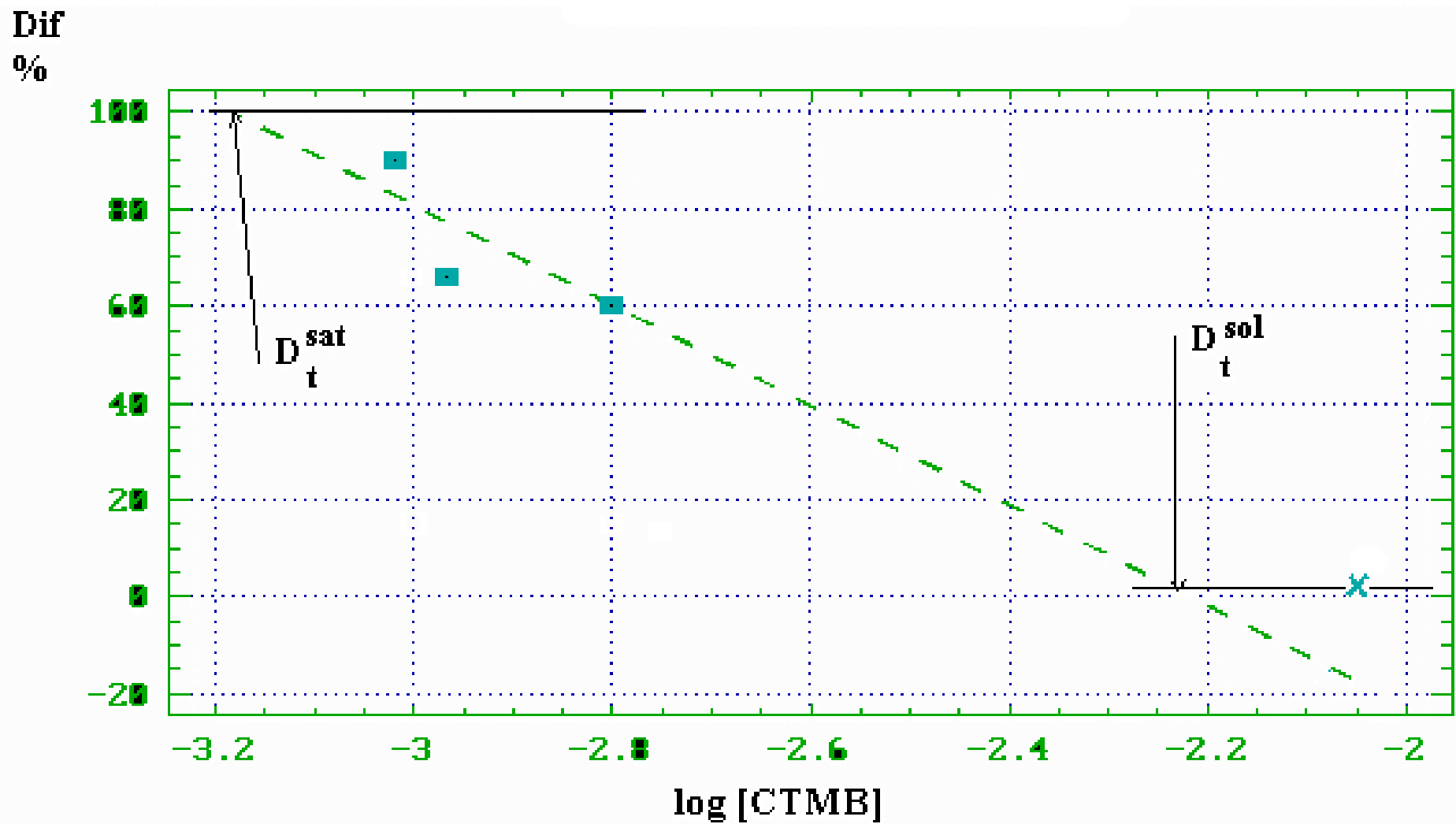

**Figure 17.** *The graphic representation of $D_t^{sol}$ dependence on EYL concentration, using CTMB as surfactant. The $R_e^{sol}$ value was calculated from the $D_t^{sol}$ straight line slope.*

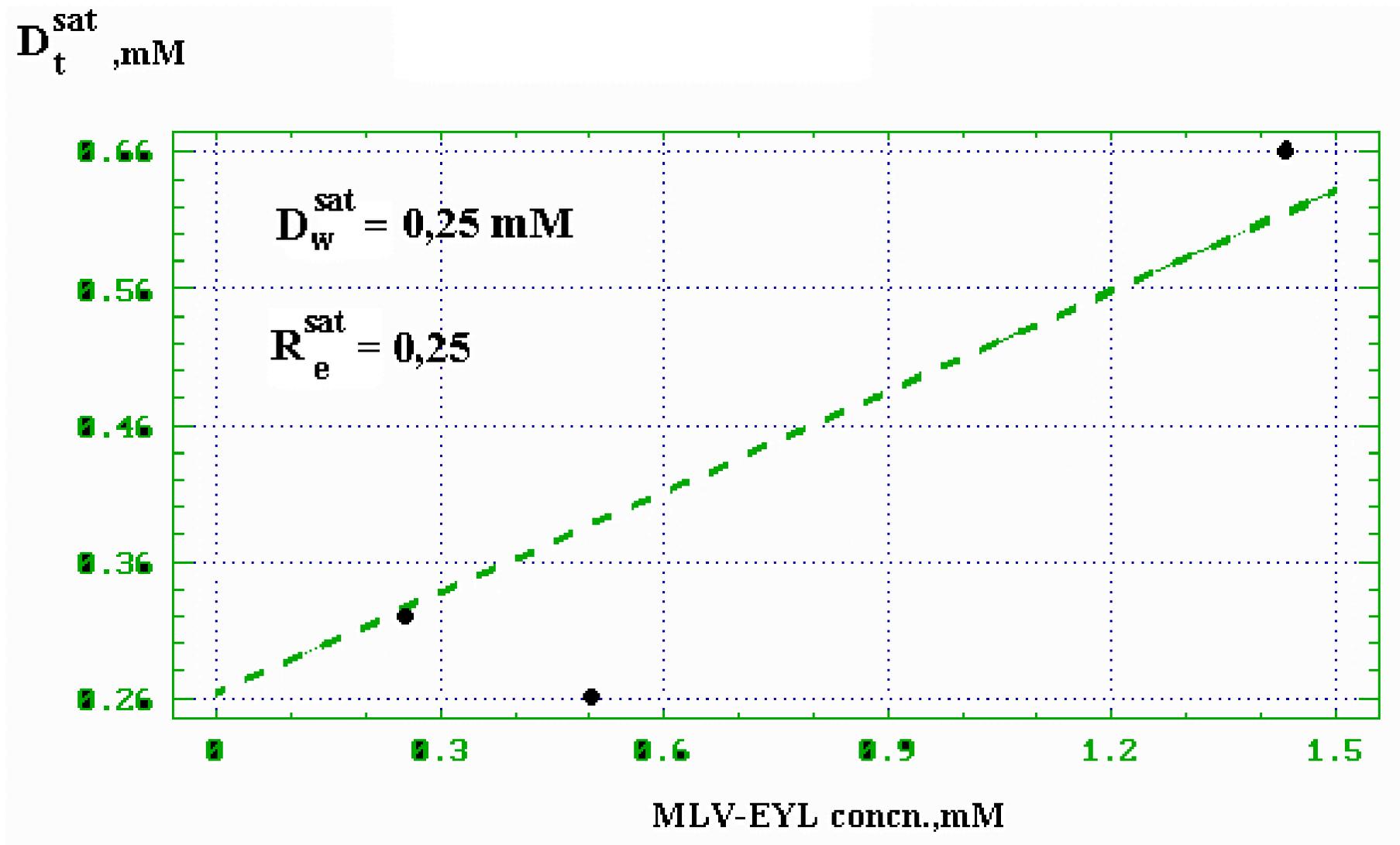

**Figure 17A.** *The graphic representation of $D_t^{sol}$ dependence on EYL concentration, using CTMB as surfactant. The $R_e^{sol}$ value was calculated from the $D_t^{sol}$ straight line slope.*

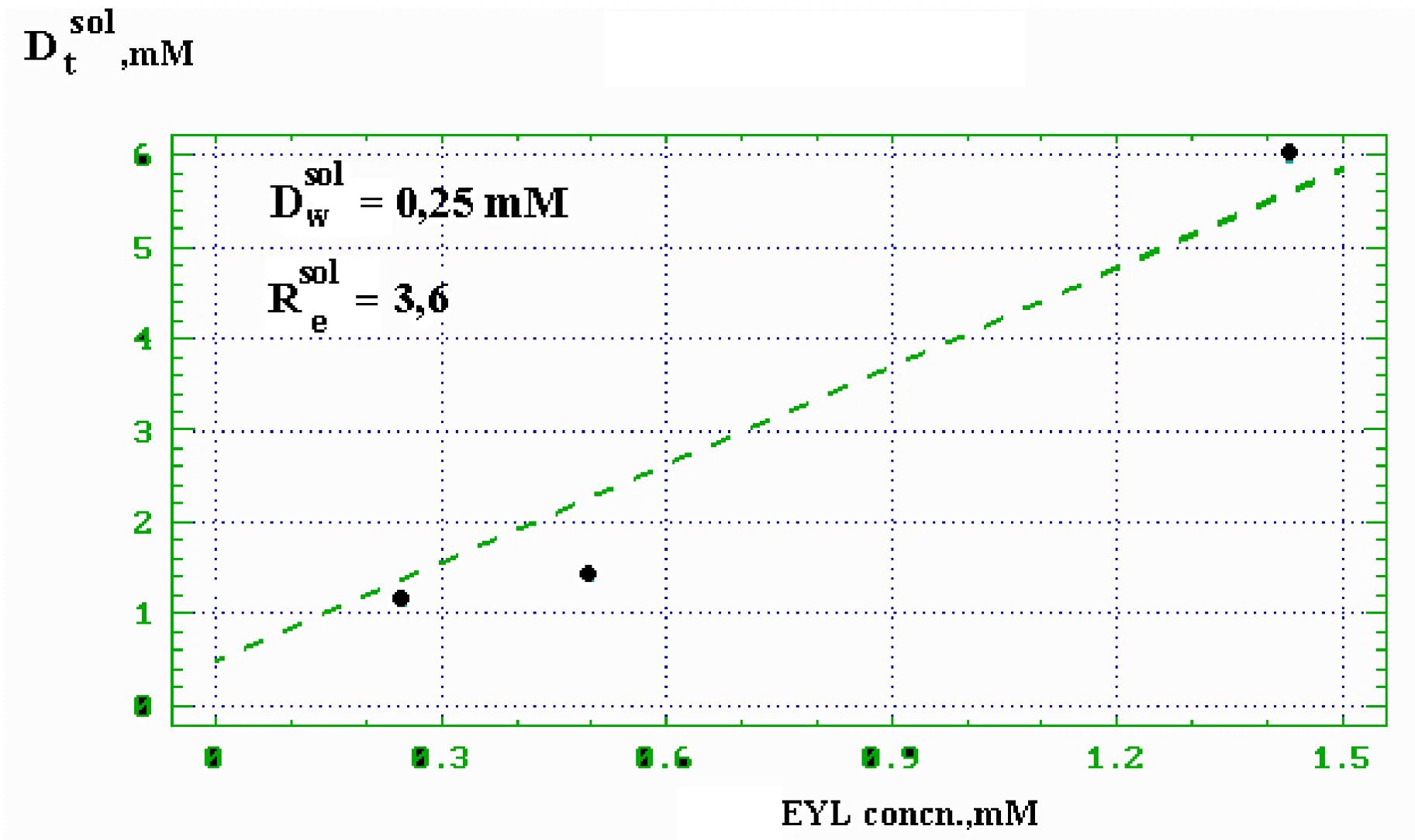